\documentclass[english,twocolumn,transaction]{IEEEtran}
\usepackage[T1]{fontenc}
\usepackage{float}
\usepackage{amsmath}
\usepackage{amsthm}
\usepackage{amssymb}
\usepackage{stackrel}
\usepackage{graphicx}
\usepackage{setspace}
\usepackage{color}
\usepackage{url}

\makeatletter

\providecommand{\tabularnewline}{\\}
\floatstyle{ruled}
\newfloat{algorithm}{tbp}{loa}
\providecommand{\algorithmname}{Algorithm}
\floatname{algorithm}{\protect\algorithmname}

\theoremstyle{plain}

\theoremstyle{definition}

\theoremstyle{plain}

\theoremstyle{plain}

\IEEEoverridecommandlockouts
\usepackage{ifpdf}
\usepackage{cite}
\ifCLASSOPTIONcompsoc
\usepackage[caption=false,font=normalsize,labelfont=sf,textfont=sf]{subfig}
\else
\usepackage[caption=false,font=footnotesize]{subfig}
\fi
\usepackage{amsmath}

\@ifundefined{showcaptionsetup}{}{
\PassOptionsToPackage{caption=false}{subfig}}
\usepackage{subfig}
\makeatother

\usepackage{babel}
\newtheorem{define}{Definition}

\newtheorem{theo}{Theorem}

\renewcommand\figurename{Fig.}
\begin{document}
\title{Economical Caching for Scalable Videos in Cache-enabled Heterogeneous Networks}

\author{Xuewei Zhang, ~\IEEEmembership{Student Member,~IEEE}, Tiejun Lv, ~\IEEEmembership{Senior Member,~IEEE},\\
Yuan Ren, Wei Ni, ~\IEEEmembership{Senior Member,~IEEE},  Norman C. Beaulieu, \\~\IEEEmembership{Fellow,~IEEE},
and Y. Jay Guo, ~\IEEEmembership{Fellow,~IEEE}
\thanks{
The financial support of the National Natural Science Foundation of
China (NSFC) (Grant No. 61671072 and 61801382) is gratefully acknowledged. (\emph{Corresponding author: Tiejun Lv.})

X. Zhang, T. Lv and N. C. Beaulieu are with the School of Information and Communication
Engineering, Beijing University of Posts and Telecommunications (BUPT), Beijing
100876, China (e-mail: \{zhangxw, lvtiejun, nborm\}@bupt.edu.cn).

Yuan Ren is with the School of Communication and Information
Engineering, Xi'an University of Posts and Telecommunications (XUPT), Xi'an
710121, China (e-mail: renyuan@xupt.edu.cn).

W. Ni is with Data61, Commonwealth Scientific and Industrial Research,
Sydney, NSW 2122, Australia (e-mail: Wei.Ni@data61.csiro.au).

Y. J. Guo is with the Global Big Data Technologies Centre,
University of Technology Sydney, Ultimo, NSW 2007, Australia
(e-mail: jay.guo@uts.edu.au).
}}

\maketitle

\begin{abstract}
We develop the optimal economical caching schemes in cache-enabled heterogeneous networks,
while delivering multimedia video services with personalized viewing qualities to mobile users.
By applying scalable video coding (SVC),
each video file to be requested is divided into one base layer (BL) and several enhancement layers (ELs).
In order to assign different transmission tasks,
the serving small-cell base stations (SBSs) are grouped into $K$ clusters.
The SBSs are able to cache and cooperatively transmit BL and EL contents to the user.
We analytically derive the expressions for successful transmission probability and ergodic service rate,
and then the closed form of EConomical Efficiency (ECE) is obtained.
In order to enhance the ECE performance,
we formulate the ECE optimization problems for two cases.
In the first case, with equal cache size equipped at each SBS,
the layer caching indicator is determined.
Since this problem is NP-hard, after the $l_{0}$-norm approximation,
the discrete optimization variables are relaxed to be continuous, and this relaxed problem is convex.
Next, based on the optimal solution derived from the relaxed problem,
we devise a greedy-strategy based heuristic algorithm to achieve the near-optimal layer caching indicators.
In the second case, the cache size for each SBS,
the layer size and the layer caching indicator are jointly optimized.
This problem is a mixed integer programming problem, which is more challenging.
To effectively solve this problem, the original ECE maximization problem is divided into two subproblems.
These two subproblems are iteratively solved until the original optimization problem is convergent.
Numerical results verify the correctness of theoretical derivations.
Additionally, compared to the most popular layer placement strategy,
the performance superiority of the proposed SVC-based caching schemes is testified.
\end{abstract}

\section{Introduction}
\renewcommand\figurename{Fig.}
The total amount of the data traffic generated by mobile users is increasing rapidly these days,
and it is predicted that a majority of the traffic will be spent on multimedia video services \cite{2016Ericsson}.
Notably, multimedia video services are playing an indispensable role in our daily lives.
With this data-demanding situation,
more serious requirements and challenges are imposed on backhaul links,
which deliver the required contents from the core networks to the wireless edges,
such as the serving base stations (BSs).
For large-scale video transmissions, the limited capacity of backhaul links becomes the bottleneck of the wireless networks.
In this regard, to effectively alleviate the severe backhaul burden and improve the service delay,
wireless caching is proposed,
which deserves more attraction and investigation for the Fifth Generation (5G) communication networks
\cite{poularakis2016exploiting,Tao2015Content,Lu2018Coded,Zhou2016Stochastic}.

With wireless caching, during the time periods with light traffic,
the contents predicted to be requested in peak hours are pre-delivered from the core networks via backhaul links \cite{Chen2017Cooperative}.
These contents are then locally stored in the BSs.
When the cached contents are demanded by mobile subscribers,
they can be immediately transmitted without repeated backhaul deliveries,
which is more time- and energy-efficient.
In general, depending on the file placement policies, there are two types of wireless caching,
namely, uncoded caching \cite{Zhang2017Multicast,Tao2015Content} and coded caching \cite{Xu2017Fundamental,Lampiris2018Adding}.
In the first type of caching, complete videos, especially the popular ones, are stored in the local cache of BSs.
In the second caching scheme, different segments or encoded packets of the video files are locally stored.
Among multiple coded caching schemes, the network coding-based caching policies,
such as the maximum distance separable (MDS) coding-based caching \cite{Liao2016Optimizing,Gabry2016On,Chen2015Optimal}
and the random linear network coding (RLNC)-based caching
\cite{Xu2017Modeling},  have attracted a lot of attention and interest.
On a parallel avenue, random caching has been advocated recently \cite{Wen2017Random, cui2016analysis},
in which content files or their random combinations are selected and stored with the optimal caching probabilities.
Owing to the superb advantages of wireless caching,
it has been extensively investigated in numerous scenarios,
such as cloud radio access networks (C-RAN) \cite{Tao2015Content,Dai2018Optimized},
small cell networks \cite{tamoor2016caching,Zhang2018Near}, heterogeneous networks \cite{Zhang2017Multicast,Zhang2018Energy},
and device-to-device communications \cite{chen2016cache,Deng2018The}.
Wireless caching is promising to alleviate the serious resource waste and traffic burden of backhaul links.
Additionally, it has the ability to enhance the system performance in terms of service delay and content hit probability.

Nowadays, there are a considerable number of multimedia video services,
and different viewing qualities are required for different types of services.
Generally speaking, mobile subscribers usually prefer basic perceptual qualities for sports games and news reports,
while high definition viewing qualities are required for entertainment shows and movies.
To shed light on this, scalable video coding (SVC) has been developed from the traditional advanced video coding (AVC) \cite{Schwarz2007Overview}.
SVC is capable of flexibly and conveniently deleting or adding parts of the bit streams
to adapt to diverse personalized user demands and dynamic network environments.
By employing SVC in wireless caching, each video file to be transmitted is divided into two parts,
namely, a base layer (BL) and several enhancement layers (ELs).
It should be pointed out that different content layers are assigned with different importance levels for decoding.
In specific, BL content is the fundamental part for a scalable video,
and can provide the most basic viewing quality.
Supplementing ELs to the BL will provide high definition viewing quality \cite{Ostovari2015Scalable},
and more ELs lead to more enhanced perceptual experience.
Recently, some researchers have been concentrating on combining SVC with wireless caching.
In particular, Zhan \emph{et al}. in \cite{ZhanContent} designed an effective SVC-based layer placement scheme,
which was able to largely reduce the content download time.
In literature \cite{Xiang2018Cache},
each video file was encoded by SVC.
In case of the untrusted cache helpers, the BL was securely transmitted.
By designing the optimal transmit beamforming vectors and cache placement strategy,
the minimal transmit power consumption was obtained.
In our early work \cite{Zhang2018Near}, by proposing the near-optimal layer placement algorithm,
we maximized the average offloaded traffic to relieve the severe traffic burden of the macro-cell BS (MBS) and reduce the service delay.

The aforementioned researches concentrated on analyzing the successful transmission rate,
average content download time, average offloaded traffic and etc.
However, these metrics cannot fully evaluate and exploit the performance gain of combining SVC with wireless caching.
This is because these researches only intended to enhance the system performance,
while the huge resource cost was not taken into consideration.
To this end, EConomical Efficiency (ECE) is recommended as an important design metric
\cite{Peng2017Cost,Zhang2018Energy2,Zhang2017Cost},
which is able to balance the network revenue and resource cost.
The ECE has been studied in the literature.
Specifically, in \cite{Peng2017Cost}, Peng \emph{et al.} investigated the economical energy efficiency (EE) in C-RAN,
where the conventional EE and the cost of wired/wireless fronthaul links were jointly considered.
In \cite{Zhang2018Energy2}, the system metric of ECE was used as a tool to
proclaim basic findings and then boost the trade-off between EE and spectral efficiency (SE).
Additionally, Zhang \emph{et al.} in \cite{Zhang2017Cost} devised
the cost-effective content placement scheme in heterogeneous networks.
Under the limited cache size, the authors maximized the network capacity.
However, all of the studies related to ECE have overlooked the issue of different perceptual qualities,
and more performance gain could be obtained if the ECE and SVC are combined into the cache-enabled networks.

In this paper, we investigate the optimal economical caching schemes,
aiming to optimize the ECE of cache-enabled heterogeneous networks.
In the meanwhile, personalized viewing experiences of mobile users are taken into account.
Encoded by SVC, each video file to be transmitted is divided into one BL and several ELs.
The user with basic viewing quality requirement is provided with the standard definition video (SDV),
and only the BL content is delivered.
Otherwise, both BL and EL contents are jointly transmitted to offer the high definition video (HDV),
so that the user can enjoy excellent perceptual experience.
In the network of interest, the serving small-cell BSs (SBSs) are divided into $K$ clusters.
These SBSs are able to cache the SVC-based content layers,
and then cooperatively transmit cached layers to users when requested.
If some required layers cannot be exploited in the local cache of SBSs,
the nearest MBS will retrieve the missing layers from the core networks.
We analytically derive the successful transmission probabilities (STPs) and ergodic service rates (ESRs),
and then the closed-form expression for ECE is obtained.
To enhance the ECE performance,
under the limited cache size, we formulate the ECE optimization problems for two cases.
In the first case, equal cache size is equipped at each SBS,
and the layer caching indicator is optimized.
In the second case, we consider the more practical scenario
where the cache size for each SBS, the layer size and the layer caching indicator are jointly optimized.
The main contributions of our paper are summarized in the following:
\begin{itemize}
\item In this paper, we consider the different perceptual requirements of multimedia users in cache-enabled heterogeneous networks,
which has yet to be studied in the literature.
In specific, if the basic viewing quality is demanded by the user,
only BL of the required content is delivered;
otherwise, both BL and EL contents are jointly transmitted to the user to provide HDV service.
\item In the cases where the required content layers are provided by cooperative SBSs and the nearest MBS,
we derive the expressions for STP and ESR by applying stochastic geometry.
Accordingly, the closed-form ECE is obtained to balance the network revenue and resource cost.
To the best of our knowledge,
this is the first time to consider the ECE and different viewing requirements of uses in wireless caching.
\item When equal cache size is available at each SBS,
an NP-hard, integer programming, ECE maximization problem is formulated.
We convexify the problem by taking the $l_{0}$-norm approximation
and relaxing the discrete optimization variables to be continuous,
and finally solve the relaxed problem efficiently with convex techniques.
Based on the solution, we devise a greedy-strategy based algorithm to acquire the near-optimal layer caching indicators.
The sub-optimality property of this algorithm is also analyzed.
\item We further consider the practically increasing scenario
where the layer caching indicator, the cache size for each SBS and the layer size are jointly optimized.
This is a more challenging mixed integer programming problem.
To effectively solve it, we divide the original ECE maximization problem
 into a layer placement subproblem and a cache size and layer size allocation subproblem.
The first subproblem can be solved by taking the $l_{0}$-norm approximation and the proposed greedy-strategy based algorithm.
The second subproblem is convex, and can be readily solved.
The two subproblems iterate until convergence.
\end{itemize}

$\;$The remaining parts of this paper are organized as follows.
In Section II, we introduce the network model, the SVC-based layer caching scheme as well as the power consumption model of the considered network.
In Section III, the STP and ESR expressions are analytically derived, and then the system metric ECE is obtained.
In Section IV, we formulate and solve the ECE maximization problem with equal cache size.
The ECE maximization problem with unequal cache size and layer size is formulated and solved in Section V.
Extensive numerical results are presented in Section VI, and this paper concludes in Section VII.
\section{System Model}
In this section, we introduce the network model, the SVC-based layer caching
scheme as well as the power consumption model of the observed network.
\subsection{Network Model}
The heterogeneous network with two tiers, i.e., the MBS tire and the SBS tier, is considered in this paper.
In the observed network, the signal-antenna MBSs and SBSs are independently and identically distributed,
and the locations of these BSs obey two homogeneous Poisson Point Processes (PPPs),
denoted by $\mathrm{\Phi_{m}}$ and $\mathrm{\Phi_{s}}$.
The density of the two processes are $\lambda_{\rm{m}}$ and $\lambda_{\rm{s}}$, respectively.
Without loss of any generality, we concentrate on the analysis of a typical user,
whose location is in the center of the observed two-tire heterogeneous network
\cite{Wen2017Random,zhang2016cache,Chen2017Cooperative,Huang2014Energy}.\footnote{
In this paper, we consider the network from the user centric perspective,
and the positions of the surrounding serving and interfering BSs of a typical user follow PPPs.
The performance analysis of each of the other users would remain the same,
by setting that user as the center of the observed network.
The user centric model can decouple the analysis of different users,
and provide analytical tractability and illustration convenience.}
\footnote{The mobility of users is not explicitly taken into account in this work.
Nevertheless, the mobility is implicitly captured under the statistical nature of the ECE analysis.
This is because, given a realization of the PPP,
a random latitudinal shift of the realization can be a new valid realization of the PPP.}
To execute different transmission tasks, the serving SBSs are divided into $K$ clusters.
In the circular area with radius $d_{1}$, the deployed SBSs form cluster $\mathcal{S}_{1}$,
and the number of SBSs belonging to this cluster is ${S}_{1}=|\mathcal{S}_{1}|$.
The other SBSs located in the annular area with radii $d_{k-1}$
and $d_{k}$ ($d_{k-1}<d_{k}$, $k=2,3,...,K$) form cluster $\mathcal{S}_{k}$,
and the number of SBSs in cluster $\mathcal{S}_{k}$ is ${S}_{k}=|\mathcal{S}_{k}|$.
All SBSs are sorted in the ascending order of the distances between the user and SBSs.

\begin{figure}[t]
\centering{}\includegraphics[scale=0.37]{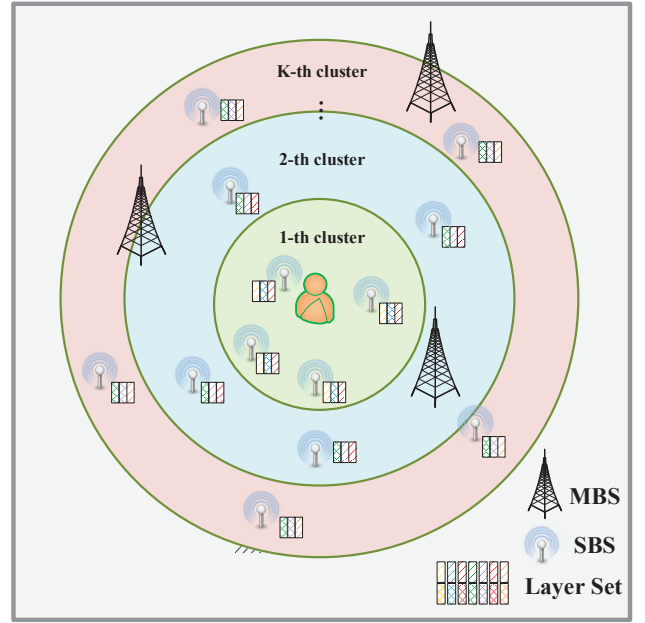}
\caption{In the proposed system model,
a typical user is located at the center of the observed network,
and the serving SBSs are grouped into $K$ clusters.
As an example, each video file is divided into two layers.}
\label{System_Model}
\end{figure}

With the help of SVC, each video file to be requested and transmitted is divided into $L$ layers.
Let $l=1$ denote the index for the BL content, and $l=2,...,L$ indicate the ELs.
The BL is expected to offer the most fundamental perceptual quality,
and the user with EL contents can obtain superior video quality.
The video file with BL content can provide the basic viewing quality,
and this kind of video is defined as SDV.
On the other hand, the HDV contains both BL and ELs,
and is able to provide superior perceptual experiences to mobile users.
The SBSs are assigned to locally cache the SVC-based video layers,
and then forward them to the user.
Note that SBSs belonging to the same cluster are expected to cache and transmit the same copies of required content layers.
This scheme can increase the transmission diversity \cite{Chen2017Cooperative} and enhance the received signal strength,
such that the STP for delivering the required layers can be improved.
When the serving SBSs cannot provide required layers to the user,
the nearest MBS will proceed to deliver uncached video layers from the core networks through backhaul links.
It is worthwhile to notice that the line-of-sight (LoS) propagation plays the most important role in short-range communications.
Thus, it is reasonably assumed that the distance between the serving SBS and the user
has the strongest impact on the received signal strength \cite{Xu2017Modeling,Wildemeersch2014Successive},
which means that closer SBS provides larger signal strength.
Based on this assumption, when adopting successive interference cancellation (SIC) for signal decoding,
the transmitted data symbols from SBSs deployed in cluster $\mathcal{S}_{1}$ can be firstly decoded.
If this signal is obtained without error, it can be removed from the original received signal.
Next, the signals from clusters $\mathcal{S}_{k}$ ($k=2,...,K$) will be subsequently decoded.
Besides, the coexisting MBSs and SBSs are allocated with orthogonal spectrum bands,
such that the inter-tier interference can be suppressed.
A detailed example of the considered system model is illustrated in Fig. \ref{System_Model}.

If the required content layers are cached in the SBSs belonging to cluster $\mathcal{S}_{k}$,
these layers will be directly transmitted to the typical user.
The received signal from cluster $\mathcal{S}_{k}$ at the typical user's side can be written as
\begin{align}
y_{k}=\sum_{l\in\mathcal{S}_{k}}h_{l}r_{l}^{-\frac{\alpha_{\rm{s}}}{2}}t_{l}\sqrt{P_{\rm{s}}}+\sum_{s\in\mathrm{\Phi_{\rm{s}}}\setminus\mathcal{S}_{k}}h_{s}r_{s}^{-\frac{\alpha_{\rm{s}}}{2}}t_{s}\sqrt{P_{\rm{s}}}+z,\label{signal1}
\end{align}
\noindent where $P_{\rm{s}}$ and $\alpha_{\rm{s}}$ are the transmit power and path loss exponent of SBSs, respectively;
$h_{l}$ represents the channel gain from the $l$-th SBS,
following the complex Gaussian distribution with zero mean and unit variance, i.e., $\mathcal{CN}(0,1)$;
$r_{l}$ denotes the distance between the user and the $l$-th SBS;
$t_{l}$ is the transmitted data symbol from the $l$-th SBS, satisfying $\mathrm{\mathbb{E}}[t_{l}^{2}]=1$;
and $z$ refers to the additive white Gaussian noise.
After SIC, the received signal $y_{k}$ can be written as
\begin{align}
y_{k}=\sum_{l\in\mathcal{S}_{k}}h_{l}r_{l}^{-\frac{\alpha_{\rm{s}}}{2}}t_{l}\sqrt{P_{\rm{s}}}+\sum_{s\in\mathrm{\Phi_{\rm{s}}}\setminus\mathcal{S}_{1},...,\mathcal{S}_{k}}h_{s}r_{s}^{-\frac{\alpha_{\rm{s}}}{2}}t_{s}\sqrt{P_{\rm{s}}}+z,
\end{align}
\noindent where the interference generated by the SBSs closer than
those located in cluster $S_{k}$ is successfully cancelled.

On the other hand, when some of the required video layers cannot be obtained from the local cache of SBSs,
the nearest MBS is supposed to deliver the missing layers.
The received signal from the nearest MBS, denoted by $m_{0}$,
can be given by
\begin{align}
&y_{m_{0}}=\nonumber\\
&h_{m_{0}}r_{m_{0}}^{-\frac{\alpha_{\rm{m}}}{2}}q_{m_{0}}\sqrt{P_{\rm{m}}}+\sum_{m\in\mathrm{\Phi_{\rm{m}}}\setminus m_{0}}h_{m}r_{m}^{-\frac{\alpha_{\rm{m}}}{2}}q_{m}\sqrt{P_{\rm{m}}}+z,
\end{align}

\noindent where $P_{\rm{m}}$ and $\alpha_{\rm{m}}$ are the transmit power and path loss exponent of MBSs, respectively;
$h_{m_{0}}$ and $h_{m}$ represent the channel gains from the nearest MBS and other MBSs,
following $\mathcal{CN}(0,1)$;
$r_{m_{0}}$ denotes the distance between the user and the nearest MBS;
$r_{m}$ denotes the distance between the user and other MBSs;
moreover, $q_{m_{0}}$ and $q_{m}$ refer to the transmitted data symbols from the nearest MBS and other MBSs,
satisfying $\mathrm{\mathbb{E}}[q_{m_{0}}^{2}]=\mathrm{\mathbb{E}}[q_{m}^{2}]=1$.

It is a common phenomenon that the interference generated by numerous coexisting BSs
has more negative impact than the noise in wireless networks.
To this end, the interference-limited network is taken into account,
where the background noise at the user's side is assumed to be negligible.
As a result, the received signal-to-interference ratio (SIR) from cluster $\mathcal{S}_{k}$ is given by
\begin{align}
&\mathrm{SIR}_{k}
 =\frac{\left|\sum_{l=S_{k-1}+1}^{S_{k}}h_{l}r_{l}^{-\frac{\alpha_{\rm{s}}}{2}}\right|^{2}}{\sum_{s\in\mathrm{\Phi_{\rm{s}}}\mathrm{\setminus}\mathcal{S}_{1},...,\mathcal{S}_{k}}\left|h_{s}\right|^{2}r_{s}^{-\alpha_{\rm{s}}}}.
\end{align}
\noindent The SIR from the nearest MBS is written as
\begin{align}
\mathrm{SIR}_{m_{0}}=\frac{\left|h_{m_{0}}\right|^{2}r_{m_{0}}^{-\alpha_{\rm{m}}}}{\sum_{m\in\mathrm{\Phi_{\rm{m}}\setminus}m_{0}}\left|h_{m}\right|^{2}r_{m}^{-\alpha_{\rm{m}}}}.
\end{align}
\subsection{SVC-based Layer Caching Scheme}
In the network of interest, the total cache size of the entire network is $M$.
The cache size allocated to each cluster is denoted by $M_{k}$,
and it satisfies $\sum_{k=1}^{K}M_{k}\leq M$.
Considering the same copies of required content layers are cached in SBSs deployed in the same cluster,
the allocated cache size for each SBS in cluster $\mathcal{S}_{k}$ can be given by $M_{k}/S_{k}$.
There are $F$ video contents required by the user at the beginning of each transmission slot.
The maximum size of each file is $C_{f}$ ($f=1,2,...,F$).
Moreover, the size of the $l$-th layer for the $f$-th video file is denoted by $C_{f,l}$,
satisfying $\sum_{l=1}^{L}C_{f,l}\leq C_{f}$.
We sort all video files in the descending popularity order.
In such an order, more popular videos are related to smaller indices.
Assume that the video popularity is known in advance.\footnote{
The video popularity changes between different realizations of the Zipf's law,
and it is typically updated at a large time scale \cite{Liu2016Energy}.
During the time period with unchanged popularity,
by taking all cases of the random user requests into account,
we design the optimal layer placement scheme to maximize the ECE.
The same analysis and optimization can be applied again
after the popularity is updated.}
The Zipf's law is a common distribution for user requests \cite{breslau1999web}.
Following this distribution, the request probability of video files is given by \cite{shanmugam2013femtocaching}
\begin{equation}
p(f)=\frac{f^{-\alpha}}{\sum_{n=1}^{F}n^{-\alpha}},\ f=1,2,...,F, \label{eq:c-1-1}
\end{equation}
\noindent where $\alpha$ is defined as the skewness parameter to capture the request concentration degree \cite{breslau1999web}.
Typically, a larger value of $\alpha$ manifests that fewer video contents can satisfy the majority of user requests.
Apart from the request probability of these video files, the quality preference needs to be characterized as well.
The preference for SDV of the $f$-th file is given by \cite{Wu2016Caching}
\begin{align}
g_{\rm{SDV}}(f)=\frac{f-1}{F-1}.
\end{align}
\noindent Accordingly, the preference for HDV is $g_{\rm{HDV}}(f)=1-g_{\rm{SDV}}(f)=\frac{F-f}{F-1}$.
When HDV is demanded by the user, it is supposed that all ELs possess the same popularity.
Hence, the request probability for the divided content layers can be calculated as
\begin{equation}
p_{f}(l)=\begin{cases}
p(f)\cdotp\frac{f-1}{F-1}, & l=1,\\
p(f)\cdotp\frac{F-f}{(F-1)(L-1)}, & l=2,...,L.
\end{cases}
\end{equation}

Employing SVC, each video file to be requested is divided into several layers.
Due to the limited cache size and power resource of the observed network,
it is critical to determine which layer needs to be placed in the local cache of the SBSs
to achieve the optimal layer placement.
Thus, we define the binary variable $x_{f,l,k}$ as the layer caching indicator,
which is used to judge whether the  $l$-th layer of the $f$-th video is cached in the SBSs deployed in
cluster $\mathcal{S}_{k}$.
If $x_{f,l,k}=1$, the $l$-th layer of the $f$-th video is locally stored in SBSs belonging to $\mathcal{S}_{k}$.
If $x_{f,l,k}=0$, this video layer cannot be locally obtained.
For notational simplicity, all layer caching indicators are collected in a matrix $\mathrm{\mathbf{x}}\in\mathbb{R^{\mathrm{F\times L\times K}}}$.
\subsection{Power Consumption Model}
In the considered cache-enabled networks, the total power consumption consists of four parts,
namely, the transmit power consumption $P_{\rm{T}}$, the caching power consumption $P_{\rm{ca}}$,
the backhaul delivery consumption $P_{\rm{bh}}$ and the fixed power consumption $P_{\rm{fix}}$.
Then, the total power consumption can be modeled by
\begin{align}
P_{\rm{Total}}=P_{\rm{T}}+P_{\rm{ca}}+P_{\rm{bh}}+P_{\rm{fix}}.\label{Total_Power}
\end{align}
\noindent The details for these power consumptions are provided as follows.

When cached video layers are requested,
the serving SBSs are required to cooperatively transmit them to the typical user.
These SBSs are in the active mode,
and the power consumptions for content transmission and some fixed budgets do exist.
Note that the MBS is always active so as to guarantee the basic converge
and transmit the necessary signalings to its serving users.
Then, the transmit power consumption $P_{\rm{T}}$ can be calculated as
\begin{align}
P_{\rm{T}}=\sum_{k=1}^{K}\left\Vert \sum_{f=1}^{F}\sum_{l=1}^{L}p_{f}(l)x_{f,l,k}\right\Vert _{0}S_{k}\zeta_{\rm{s}}P_{\rm{s}}+\zeta_{\rm{m}}P_{\rm{m}},\label{tr_Power}
\end{align}
\noindent where $\zeta_{\rm{s}}$ and $\zeta_{\rm{m}}$ denote the power inefficiencies of the power amplifiers for SBS and MBS, respectively.
The $l_{0}$-norm in (\ref{tr_Power}) is used to determine
whether the files cached in SBSs belonging to cluster $\mathcal{S}_{k}$ are required by the user.
In the similar manner, the fixed power consumption $P_{\rm{fix}}$,
caused by site-cooling, managing circuit components and running the oscillator, can be given by
\begin{align}
P_{\rm{fix}}=\sum_{k=1}^{K}\left\Vert \sum_{f=1}^{F}\sum_{l=1}^{L}p_{f}(l)x_{f,l,k}\right\Vert _{0}S_{k}P_{\rm{fix,s}}+P_{\rm{fix,m}},\label{fix_Power}
\end{align}
\noindent where $P_{\rm{fix,s}}$ and $P_{\rm{fix,m}}$ are the fixed power consumptions for SBS and MBS, respectively.

As for the caching power consumption $P_{\rm{ca}}$ and backhaul delivery consumption $P_{\rm{bh}}$,
the energy-proportional model has been widely adopted in the content-centric networks \cite{Liu2016Energy, Li2013Energy}.
As described in \cite{Gabry2016On,Choi2012In},
the caching power consumption is proportional to the total number of the data bits
that are placed in the local cache of BSs.
Similarly, the backhaul power consumption scales in proportion to the total number of the data bits
that are delivered from the core networks over backhaul links.
Following this energy-proportional model, $P_{\rm{ca}}$ and $P_{\rm{bh}}$ can be calculated as
\begin{align}
&P_{\rm{ca}}=\sum_{k=1}^{K}\sum_{f=1}^{F}\sum_{l=1}^{L}x_{f,l,k}c_{\rm{ca}}C_{f,l}S_{k},\label{ca_Power}\\
& P_{\rm{bh}}=\sum_{k=1}^{K}\sum_{f=1}^{F}\sum_{l=1}^{L}p_{f}(l)(1-x_{f,l,k})c_{\rm{bh}}C_{f,l},\label{bh_Power}
\end{align}
\noindent where $c_{\rm{ca}}$ and $c_{\rm{bh}}$ are the caching and backhaul consumption coefficients, respectively.
Substituting (\ref{tr_Power}), (\ref{fix_Power}), (\ref{ca_Power}) and (\ref{bh_Power}) into (\ref{Total_Power}),
the expression for total power consumption $P_{\rm{Total}}$ can be obtained.
\section{The Analysis and Derivations of ECE}
In this section, we give the definition of ECE,
and derive the expressions for STP and ESR.
Then, the closed-form ECE can be developed.
\subsection{The Analysis of ECE}
We intend to consider the economical gain of the proposed SVC-based layer caching scheme.
To this end, the system metric ECE is defined to measure the net profit of the network at every second, as given by \cite{Zhang2018Energy2}
\begin{gather}
\mathrm{ECE}=Re-C,\label{ECE}
\end{gather}
\noindent where $Re$ is the network revenue and $C$ is the total cost.
The ECE is a more comprehensive metric to balance the network revenue and total cost.

First, let us discuss the revenue of wireless networks.
The network revenue represents the economical gain at every second
when providing data services to multimedia users.
Typically, different kinds of mobile services possess different revenue functions,
where the linear, logarithmic and constant functions are adopted as revenue functions
for services with voice traffic, limited-volume traffic and unlimited-volume traffic,
respectively \cite{Patcharamaneepakorn2016Spectral}.
The data traffic generated by multimedia video services can be regarded as the volume-limited traffic,
thus the logarithmic function is employed as the revenue function,
as given by \cite{Zhang2018Energy2}
\begin{align}
{\rm{Revenue}}\overset{\Delta}{=}k_{\rm{r}}R_{\rm{ref}}\log_{2}(1+\frac{R}{R_{\rm{ref}}}).\label{Re}
\end{align}
\noindent In (\ref{Re}), $k_{\rm{r}}$ is the unit price per data bit, and $R$ is the service rate.
In addition, $R_{\rm{ref}}$ represents the referenced date rate,
which is able to guarantee the minimum quality of service (QoS) requirement for the user.
As mentioned earlier, the total revenue of the proposed SVC-based layer caching scheme can be given by
\begin{align}
Re= & k_{\rm{r}}\sum_{k=1}^{K}\sum_{f=1}^{F}\sum_{l=1}^{L}p_{f}(l)[x_{f,l,k}R_{\rm{ref},\rm{s}}\log_{2}(1+\frac{R_{k}}{R_{\rm{ref},\rm{s}}})\nonumber\\
 & +(1-x_{f,l,k})R_{\rm{ref},\rm{m}}\log_{2}(1+\frac{R_{m_{0}}}{R_{\rm{ref},\rm{m}}})]\nonumber\\
= & k_{\rm{r}}\sum_{k=1}^{K}\sum_{f=1}^{F}\sum_{l=1}^{L}p_{f}(l)[x_{f,l,k}(R_{\rm{ref},\rm{s}}\log_{2}(1+\frac{R_{k}}{R_{\rm{ref},\rm{s}}})\nonumber \\ &-R_{\rm{ref},\rm{m}}\log_{2}(1+\frac{R_{m_{0}}}{R_{\rm{ref},\rm{m}}}))]+R_{0},\label{eq:Revene}
\end{align}
\noindent where
\begin{align}
R_{0}=Kk_{\rm{r}}R_{\rm{ref},\rm{m}}\log_{2}(1+\frac{R_{m_{0}}}{R_{\rm{ref},\rm{m}}}).
\end{align}
In (\ref{eq:Revene}),
$R_{\rm{ref},\rm{s}}=W_{\rm{s}}\log_{2}(1+\gamma_{\rm{s}})$
and $R_{\rm{ref},\rm{m}}=W_{\rm{m}}\log_{2}(1+\gamma_{\rm{m}})$, where
$W_{\rm{s}}$ and $W_{\rm{m}}$ are the allocated system bandwidths for SBSs and MBSs,
and $\gamma_{\rm{s}}$ and $\gamma_{\rm{m}}$
are set as the minimum QoS requirements for cooperative SBSs and the nearest MBS.
Additionally, $R_{k}$ and $R_{m_{0}}$ are denoted as the ESRs achieved by SBSs in cluster $\mathcal{S}_{k}$ and the nearest MBS,
whose definitions are given as follows.
\begin{define}
The ESRs that are contributed by SBSs deployed in the $k$-th cluster and the nearest MBS are defined as
\begin{align}
&R_{k}=W_{\rm{s}}\mathrm{\mathbb{E}}\left\{\log_{2}(1+\mathrm{SIR}_{k})|\mathrm{SIR}_{k}\geq\gamma_{\rm{s}}\right\},\label{SBS_Rate_1}\\
&R_{m_{0}}=W_{\rm{m}}\mathrm{\mathbb{E}}\left\{\log_{2}(1+\mathrm{SIR}_{m_{0}})|\mathrm{SIR}_{m_{0}}\geq\gamma_{\rm{m}}\right\} \label{MBS_Rate_1},
\end{align}
\noindent where the expectations are related to the small-scale fading,
as well as PPP-distributions of the locations of SBSs and MBSs.
\end{define}

In the unit time of one second, the total cost is the monetary overhead when running the cache-enabled network.
The cost is proportional to the total power consumption, as given by
\begin{align}
  C&=k_{\rm{c}}P_{\rm{Total}}\nonumber\\
  &=k_{\rm{c}}[\sum_{k=1}^{K}\left\Vert \sum_{f=1}^{F}\sum_{l=1}^{L}p_{f}(l)x_{f,l,k}\right\Vert _{0}S_{k}(\zeta_{\rm{s}}P_{\rm{s}}+P_{\rm{fix,s}})\nonumber
  \end{align}
  \begin{align}
 &+\sum_{k=1}^{K}\sum_{f=1}^{F}\sum_{l=1}^{L}x_{f,l,k}C_{f,l}(c_{\rm{ca}}S_{k}-p_{f}(l)c_{\rm{bh}})]+C_{0},\label{cost}
\end{align}
\noindent where
\begin{align}
C_{0}=k_{\rm{c}}\big[K\sum_{f=1}^{F}\sum_{l=1}^{L}p_{f}(l)c_{\rm{bh}}C_{f,l}+\zeta_{\rm{m}}P_{\rm{m}}+P_{\rm{fix,m}}\big],
\end{align}
\noindent and $k_{\rm{c}}$ specifies the unit price per Joule.
Finally, by substituting (\ref{eq:Revene}) and (\ref{cost}) into (\ref{ECE}), the expression for ECE can be derived.
However, the ESRs appearing in (\ref{eq:Revene}) are not yet available,
and will be derived in the following subsection.
\subsection{The Derivations of ESRs}
In order to obtain the closed-form ECE, we need the expressions for ESR.
The derivations of $R_{k}$ and $R_{m_{0}}$ are revealed in the following theorem.

\begin{theo}
When the typical user is provided with the required video layers by SBSs in cluster $\mathcal{S}_{k}$ and the nearest MBS,
the ESRs can be calculated as
\begin{align}
 R_{k}=&W_{\rm{s}}\log_{2}(1+\gamma_{\rm{s}})+\frac{W_{\rm{s}}}{\ln2}\int_{d_{k-1}}^{d_{k}}...\int_{d_{k-1}}^{d_{k}}f_{\rm{s}}(\mathbf{x}_{k})\mathrm{d}\mathbf{x}_{k}\nonumber\\
 & \int_{\gamma_{\rm{s}}}^{\infty}\frac{\Pr(\mathrm{SIR}_{k}\geq t|\mathbf{r}_{k}=\mathbf{x}_{k})}{\Pr(\mathrm{SIR}_{k}\geq\gamma_{\rm{s}}|\mathbf{r}_{k}=\mathbf{x}_{k})(1+t)}\mathrm{d}t,\\
 R_{m_{0}}=&W_{\rm{m}}\log_{2}(1+\gamma_{\rm{m}})+\frac{W_{\rm{m}}}{\ln2}\int_{0}^{\infty}f_{\rm{m}}(x_{m_{0}})\mathrm{d}x_{m_{0}}\nonumber\\
 & \int_{\gamma_{\rm{m}}}^{\infty}\frac{\Pr(\mathrm{SIR}_{m_{0}}\geq t|r_{m_{0}}=x_{m_{0}})}{\Pr(\mathrm{SIR}_{m_{0}}\geq\gamma_{\rm{m}}|r_{m_{0}}=x_{m_{0}})(1+t)}\mathrm{d}t,
\end{align}
\noindent where $d_{0}=0$; $\mathbf{x}_{k}=[x_{s_{k-1}+1},...,x_{s_{k}}]$; $\mathbf{r}_{k}=[r_{s_{k-1}+1},...,r_{s_{k}}]$;
$\Pr(\mathrm{SIR}_{k}\geq\gamma_{\rm{s}}|\mathbf{r}_{k}=\mathbf{x}_{k})$ and $\Pr(\mathrm{SIR}_{m_{0}}\geq\gamma_{\rm{m}}|r_{m_{0}}=x_{m_{0}})$
are the conditional STPs for SBSs in cluster $\mathcal{S}_{k}$ and the nearest MBS, respectively.
Moreover, the probability density functions (PDFs) of the locations of serving SBSs in $\mathcal{S}_{k}$ and the nearest MBS can be expressed by
\begin{align}
&f_{\rm{s}}(\mathbf{x}_{k})=\prod_{l=S_{k-1}+1}^{S_{k}}\frac{2x_{l}}{d_{k}^{2}-d_{k-1}^{2}},\\
&f_{\rm{m}}(x_{m_{0}})=2\pi\lambda_{\rm{m}}x_{m_{0}}\exp(-\pi\lambda_{\rm{m}}x_{m_{0}}^{2}),
\end{align}
\noindent where we set $S_{0}=0$.
\end{theo}

\begin{IEEEproof}
For brevity, the proof for this theorem is suppressed.
Readers can refer to our earlier paper for more details \cite [Theorem 1] {Zhang2018Energy}.
\end{IEEEproof}

From theorem 1, we can see that, to obtain the expressions for ESR,
the STPs need to be derived.

\begin{theo}
By applying stochastic geometry,
the STPs of cooperative SBSs belonging to cluster $\mathcal{S}_{k}$ and the nearest MBS can be calculated as
\begin{align}
 &\Pr(\mathrm{SIR}_{k}\geq\gamma_{\rm{s}})=\int_{d_{k-1}}^{d_{k}}...\int_{d_{k-1}}^{d_{k}}\prod_{l=S_{k-1}+1}^{S_{k}}\frac{2x_{l}}{d_{k}^{2}-d_{k-1}^{2}}\nonumber\\
 &\exp(-\pi\lambda_{\rm{s}}(k_{1}\gamma_{\rm{s}})^{\frac{2}{\alpha_{\rm{s}}}}G_{\alpha_{\rm{s}}}(d_{k}^{2}(k_{1}\gamma_{\rm{s}})^{-\frac{2}{\alpha_{\rm{s}}}}))\mathrm{d}\mathbf{x}_{k},\\
&\Pr(\mathrm{SIR}_{m_{0}}\geq\gamma_{\rm{m}})=(\gamma_{\rm{m}}^{\frac{2}{\alpha_{\rm{m}}}}G_{\alpha_{\rm{m}}}(\gamma_{\rm{m}}^{-\frac{2}{\alpha_{\rm{m}}}})+1)^{-1},
\end{align}

\noindent where $k_{1}=\frac{1}{\sum_{l=S_{k-1}+1}^{S_{k}}x_{l}^{-\alpha_{\rm{s}}}}$ and $G_{a}(b)=\int_{b}^{\infty}\frac{1}{1+r^{\frac{a}{2}}}\mathrm{d}r$.
\end{theo}

\begin{IEEEproof}
Please refer to Appendix A.
\end{IEEEproof}

Based on the derived expressions for STP,
the ESRs can be obtained as in (\ref{SBS_Rate}) and (\ref{MBS_Rate}) shown in the top of the next page.
\begin{algorithm*}[t]
\begin{align}
 R_{k}=&W_{\rm{s}}\log_{2}(1+\gamma_{\rm{s}})+\frac{W_{\rm{s}}}{\ln2}\int_{d_{k-1}}^{d_{k}}...\int_{d_{k-1}}^{d_{k}}\prod_{l=S_{k-1}+1}^{S_{k}}\frac{2x_{l}}{d_{k}^{2}-d_{k-1}^{2}}\mathrm{d}\mathbf{x}_{k}\nonumber\\
 &\int_{\gamma_{\rm{s}}}^{\infty}\exp(-\pi\lambda_{\rm{s}}k_{1}^{\frac{2}{\alpha_{\rm{s}}}}(t^{\frac{2}{\alpha_{\rm{s}}}}G_{\alpha_{\rm{s}}}(d_{k}^{2}(k_{1}t)^{-\frac{2}{\alpha_{\rm{s}}}})-\gamma_{\rm{s}}^{\frac{2}{\alpha_{\rm{s}}}}G_{\alpha_{\rm{s}}}(d_{k}^{2}(k_{1}\gamma_{\rm{s}})^{-\frac{2}{\alpha_{\rm{s}}}})))\frac{1}{1+t}\mathrm{d}t, \label{SBS_Rate}\\
 R_{m_{0}}=&W_{\rm{m}}\log_{2}(1+\gamma_{\rm{m}})+\frac{2\pi\lambda_{\rm{m}}W_{\rm{m}}}{\ln2}\int_{0}^{\infty}x_{m_{0}}\exp(-\pi\lambda_{\rm{m}}x_{m_{0}}^{2})\mathrm{d}x_{m_{0}}\nonumber\\
 & \int_{\gamma_{\rm{m}}}^{\infty}\exp(-\pi\lambda_{\rm{m}}x_{m_{0}}^{2}(t^{\frac{2}{\alpha_{\rm{m}}}}G_{\alpha_{\rm{m}}}(t^{-\frac{2}{\alpha_{\rm{m}}}})-\gamma_{\rm{m}}^{\frac{2}{\alpha_{\rm{m}}}}G_{\alpha_{\rm{m}}}(\gamma_{\rm{m}}^{-\frac{2}{\alpha_{\rm{m}}}})))\frac{1}{1+t}\mathrm{d}t.\label{MBS_Rate}
\end{align}
\end{algorithm*}
\noindent By substituting (\ref{SBS_Rate}) and (\ref{MBS_Rate}) into (\ref{eq:Revene}), the expression for ECE is obtained.
\section{The ECE Maximization Problem with Equal Cache Size and Layer size}
In this section, we optimize the ECE when equal cache size is equipped at each SBS.
Moreover, content layers from the same video file are of equal size,
and only the layer caching indicator needs to be determined.
\subsection{The Formulated ECE Optimization Problem and the Proposed Algorithm}
When equal cache size is equipped at each SBS,
the cache size of each SBS can be calculated as $Q=\frac{M}{\sum_{k=1}^{K}S_{k}}$.
Moreover, for content layers belonging to the same video file,
the layer sizes are identical, and can be calculated as $C_{f,l}=C_{f}/L$.
To improve the ECE performance, under the limited cache size,
the ECE optimization problem is formulated as
\begin{subequations}\label{max_2}
\begin{align}
\underset{\mathbf{x}}{\mathrm{max}}\quad & \mathrm{ECE}\label{eq:2_1}\\
\mathrm{\mathrm{s.t.}}\:\quad & \sum_{k=1}^{K}x_{f,l,k}\leq1,\forall f,\forall l,\label{eq:2_2}\\
 & \sum_{f=1}^{F}\sum_{l=1}^{L}x_{f,l,k}C_{f,l},\leq Q,\forall k,\label{eq:2_3}\\
 & x_{f,l,k}\in\{0,1\},\forall f,\forall l,\forall k.\label{eq:2_4}
\end{align}
\end{subequations}
Here, constraint (\ref{eq:2_2}) indicates
that one video layer should be placed in no more than one cluster of the SBSs,
such that the layer diversity can be enhanced and the performance of content hit probability is improved.
The cache size restriction for each SBS is captured in constraint (\ref{eq:2_3}).
(\ref{eq:2_4}) specifies the feasible range of the discrete variable $x_{f,l,k}$.

Problem (\ref{max_2}) is an integer programming problem.
It is nonconvex and NP-hard \cite{ZhanContent}.
Additionally, the $l_{0}$-norm shown in the objective function (\ref{eq:2_1}) makes the problem intractable.
To solve the problem effectively, we firstly suppress the $l_{0}$-norm.
The $l_{0}$-norm of a scalar $x$ can be approximated as \cite{Peng2017Cost}
\begin{align}
& \left\Vert x\right\Vert _{0}\approx\frac{x^{(t)}}{x^{(t)}+\tau}, \label{cost_2}
\end{align}
where  the superscript ``$(t)$'' indicates the optimal value obtained in the $t$-th iteration,
and $\tau$ is a constant parameter to characterize the approximation accuracy.
Then, the total cost can be rewritten as
\begin{align}
\tilde{C}&= k_{\rm{c}}[\sum_{k=1}^{K}\frac{\sum_{f=1}^{F}\sum_{l=1}^{L}p_{f}(l)x_{f,l,k}^{(t)}}{\sum_{f=1}^{F}\sum_{l=1}^{L}p_{f}(l)x_{f,l,k}^{(t)}+\tau}S_{k}(\zeta_{s}P_{\rm{s}}+P_{\rm{fix},\rm{s}})\nonumber
\end{align}
\begin{align}
 & +\sum_{k=1}^{K}\sum_{f=1}^{F}\sum_{l=1}^{L}x_{f,l,k}^{(t)}(c_{\rm{ca}}C_{f,l}S_{k}-p_{f}(l)c_{\rm{bh}}C_{f,l})]+C_{0}.
\end{align}
\noindent After the $l_{0}$-norm approximation,
the original Problem (\ref{max_2}) is still difficult to solve due to its NP-hard property.
Then, the discrete binary variable $x_{f,l,k}$ is relaxed to be continuous.
The relaxed ECE optimization problem is written as
\begin{subequations}\label{max_3}
\begin{align}
\underset{\tilde{\mathbf{x}}}{\mathrm{max}}\quad\: & \mathrm{\widetilde{ECE}}=Re-\tilde{C}\label{eq:3_1}\\
\mathrm{\mathrm{s.t.}}\:\quad\: & \sum_{k=1}^{K}\tilde{x}{}_{f,l,k}\leq1,\forall f,\forall l,\label{eq:3_2}\\
 & \sum_{f=1}^{F}\sum_{l=1}^{L}\tilde{x}{}_{f,l,k}C_{f,l}\leq Q,\forall k,\label{eq:3_3}\\
 & 0\leq\tilde{x}_{f,l,k}\leq1,\forall f,\forall l,\forall k,\label{eq:3_4}
\end{align}
\end{subequations}

\noindent where the matrix $\tilde{\mathbf{x}}\in\mathbb{R^{\mathrm{F\times L\times K}}}$ collects all values of $\tilde{x}$.
The relaxed Problem (33) is convex,
and its optimal solution can be generated by running off-the-shelf CVX solver, i.e., SeDuMi, iteratively
until convergence.
According to the optimal solution for (\ref{max_3}),
a greedy-strategy based algorithm is proposed to achieve the near-optimal $x_{f,l,k}$,
and more details are presented in Algorithm 1.

At the beginning of Algorithm 1, we initialize all values of $x_{f,l,k}$ to be 0,
and then $z_{f,l,k}$ is calculated to evaluate the marginal economical gain
when caching the $l$-th layer of the $f$-th video file in SBSs located in cluster $\mathcal{S}_{k}$.
We set the particular $x_{f,l,k}$ to 1 first,
which possesses the same subscript with the optimal $\tilde{x}_{f,l,k}$
that is generated by (\ref{max_3}) and can provide the largest $z_{f,l,k}$.
Afterwards, if SBSs reserve enough storage space,
the other $x_{f,l,k}$s are set to 1 successively following the sorted order of $z_{f,l,k}$.
Note that the constraints in (\ref{max_2}) are reflected in Algorithm 1.
In specific, to meet constraint (\ref{eq:2_2}), we have Step 10),
which ensures that each video layer is cached at no more than one cluster of SBSs.
Step 7) makes sure that the total size of cached layers is less than or equal to the allocated cache size for each SBS,
satisfying constraint (\ref{eq:2_3}).
When Algorithm 1 is over, all layer caching indicators are examined,
and the values of these indicators are set to 1 or remain 0 otherwise,
then constraint (\ref{eq:2_4}) is satisfied.
From these steps, the near-optimal solutions for Problem (\ref{max_2}) can be achieved.

It is noted that the {\emph{preference indifference}} \cite{Zhou2015Energy} may occur in Step 4),
which means that different content layers cached in the same cluster of SBSs may possess the same value of $z_{f,l,k}$.
To deal with this issue, we establish the following principles:
\begin{itemize}
\item When file indices for these videos are different,
the layer with smaller ratio between the file size and request probability will be cached with higher priority.
This principle can improve the file diversity and thus enhance the content hit probability.\footnote{
In the special case with equal file size,
the comparison of file-size-to-request-probability ratios degrades into the comparison of request probabilities,
which indicates that the content layer from more popular videos has higher caching priority.}
\item When file indices for these layers are identical but the layer indices are different,
the layer with smaller index will be cached with higher priority.
This is due to the fact that the layer of a scalable video with smaller index is more important in the decoding process.
\end{itemize}

\begin{algorithm}[t]
\begin{enumerate}
\item Initialization: set $x_{f,l,k}=0$ for $\forall f,\forall l,\forall k$,
and set $m_{k}=Q$ for $\forall k$.
\item Solve the relaxed optimization Problem (\ref{max_3}),
and obtain the optimal solution $\tilde{x}_{f,l,k}$.
\item For $\forall f,\forall l,\forall k,$ calculate the marginal ECE, denoted by $z_{f,l,k}$, as
\begin{align*}
 & z_{f,l,k}=k_{\rm{r}}p_{f}(l)[\tilde{x}_{f,l,k}(R_{\rm{ref},\rm{s}}\log_{2}(1+\frac{R_{k}}{R_{\rm{ref},\rm{s}}})\\
 & -R_{\rm{ref},\rm{m}}\log_{2}(1+\frac{R_{m_{0}}}{R_{\rm{ref},\rm{m}}}))+R_{\rm{ref},\rm{m}}\log_{2}(1+\frac{R_{m_{0}}}{R_{\rm{ref},\rm{m}}})]\\
 & -k_{\rm{c}}[||p_{f}(l)\tilde{x}_{f,l,k}||_{0}S_{k}(\zeta_{s}P_{\rm{s}}+P_{\rm{fix,s}})\\
 & +\tilde{x}_{f,l,k}C_{f,l}(c_{\rm{ca}}S_{k}-p_{f}(l)c_{\rm{bh}})+C_{f,l}p_{f}(l)c_{\rm{bh}}],
\end{align*}
and the matrix $\mathrm{\mathbf{Z}}\in\mathbb{R^{\mathrm{F\times L\times K}}}$ is used to gather all values of $z_{f,l,k}$.
\item Sort $\tilde{x}_{f,l,k}$ in the decreasing order of $z_{f,l,k}$,
and set $i=1$.
\item \textbf{while} ($i\leq FLK$ and $\mathrm{\mathbf{Z}}$ is not empty)
\item $\:\:$find the maximum $z_{f,l,k}$;
\item $\:\:$\textbf{if} $m_{k}\geq C_{f,l}$
\item $\:\:\:\:\:\:\:x_{f,l,k}=1;$
\item $\:\:\:\:\:\:\:m_{k}=m_{k}-C_{f,l}$;
\item $\:\:\:\:\:\:$ Delete $z_{f,l,k^{'}},\forall k^{'}\neq k$ from $\mathrm{\mathbf{Z}}$;
\item $\:\:$\textbf{end if }
\item $\:\:$Delete $z_{f,l,k}$ from $\mathrm{\mathbf{Z}}$;
\item $\:\:i=i+1;$
\item \textbf{end while }
\end{enumerate}
\caption{The proposed greedy-strategy based heuristic algorithm
         for solving Problem (\ref{max_2}).}
\end{algorithm}
\subsection{Discussions of the Proposed Algorithm 1}
{\textbf{Stability and Optimality Analysis.}} In the considered layer placement problem,
multiple video layers search for the SBSs that can provide the largest economical caching gains.
The proposed optimal layer placement Problem (\ref{max_2}) is a variation of the
{\emph{many-to-one matching problem}} for the user-SBS association,
where several users search for the best SBS and then establish connections,
and the SBS can select the optimal users to serve.
For the many-to-one matching problem, the {\emph{stability} is a key performance metric.
In order to illustrate the stability property of Algorithm 1,
we give the following definition \cite{Pantisano2014Cache}.
\begin{define}
A layer-SBS association is stable if there does not exist two layers $l_{i}$ and $l_{j}$
that are cached in SBSs belonging to two clusters $n_{i}$ and $n_{j}$, $i \ne j$,
respectively, although $l_{i}$ prefers $n_{j}$ and $l_{j}$ prefers $n_{i}$.
\end{define}

\noindent From the definition of stability, the stability property of Algorithm 1 can be revealed.

\begin{theo}
After the proposed greedy-strategy based Algorithm 1,
the layer-SBS association, i.e., the layer placement, is stable.
\end{theo}
\begin{IEEEproof}
See Appendix B.
\end{IEEEproof}

Next, the sub-optimality of the proposed Algorithm 1 is given in the following theorem.

\begin{theo}
Under the case with $FLK \rightarrow \infty$,
the relaxed Problem (\ref{max_3}) is equivalent to the original maximization Problem (\ref{max_2}),
such that the optimal layer caching indicators
can be directly acquired by solving Problem (\ref{max_3}) without invoking Algorithm 1.
\end{theo}

\begin{IEEEproof}
Similar steps for proof can be found in \cite[Theorem 2]{Xiang2018Cache}.
\end{IEEEproof}

{\textbf{Computational complexity.}}
The original Problem (\ref{max_2}) cannot be solved in the polynomial time
due to its NP-hard property.
After employing Algorithm 1, the computational complexity is largely reduced.
In specific, after we derive the optimal solution for the relaxed Problem (\ref{max_3}),
in the worst case, when Algorithm 1 is over, all layer caching indicators will be examined,
then the layer caching indicators can be determined.
Therefore, the computational complexity of Algorithm 1 is ${\cal O}(FLK)$.
\begin{algorithm}[t]
\begin{enumerate}
\item  Initialization: input the feasible $\mathbf{x}$,
the accuracy threshold $\bigtriangleup$, and the maximum number of iterations $T$.
\item  Set $\delta$ as a large value such  that $\delta\gg\bigtriangleup$,
$t=0$ and $\rm{ECE}^{(t)}=0$.
\item  \textbf{while} ($\delta>\bigtriangleup$ and $t\leq T$)
\item  With given $\mathbf{x}^{*}$, solve Problem (\ref{max_5}) by running CVX solvers,
       and obtain the optimal $\mathbf{C}^{*}$ and $\mathbf{M}^{*}$;
\item  With given $\mathbf{C}^{*}$ and $\mathbf{M}^{*}$, solve Problem (\ref{max_6}) by Algorithm 1,
       and obtain the optimal $\mathbf{x}^{*}$ and $\rm{ECE}^{*}$;
\item  $\delta=|{\rm{ECE}^{*}}-{\rm{ECE}}^{(t)}|$;
\item  $t=t+1$;
\item  ${\rm{ECE}}^{(t)}={\rm{ECE}^{*}}$;
\item  \textbf{end while}
\item  Output: The optimal cache size $\mathbf{C}^{*}$,
layer size $\mathbf{M}^{*}$, layer caching indicator $\mathbf{x}^{*}$, and the corresponding $\rm{ECE}^{*}$.
\end{enumerate}
\caption{The proposed iterative algorithm for solving Problem (\ref{max_4}).}
\end{algorithm}
\section{The ECE Maximization Problem with Unequal Cache Size and Layer Size}
For the more practical scenario, the cache size and layer size are not identical for different SBSs and video layers.
In this section, we plan to search for the optimal cache size, layer size
and layer caching indicator to optimize the ECE.
The cache size of the $k$-th cluster is denoted by $M_{k}$,
and all values of $M_{k}$ are collected in a vector $\mathbf{M}\in\mathbb{R^{\mathrm{K\times 1}}}$.
Moreover, the sizes of all video layers are collected in a matrix $\mathbf{C}\in\mathbb{R^{\mathrm{F\times L}}}$.
Then, the ECE optimization problem with unequal cache size and layer size is reformulated as
\begin{subequations}\label{max_4}
\begin{align}
\underset{\mathbf{x},\mathbf{M},\mathbf{C}}{\mathrm{max}}\quad & \mathrm{ECE}\label{eq:4_1}\\
\mathrm{\mathrm{s.t.}}\:\quad & \sum_{k=1}^{K}x_{f,l,k}\leq1,\forall f,\forall l,\label{eq:4_2}\\
& \sum_{f=1}^{F}\sum_{l=1}^{L}x_{f,l,k}C_{f,l},\leq M_{k}/S_{k},\forall k,\label{eq:4_3}\\
& \sum_{k=1}^{K}M_{k}\leq M,\label{eq:4_4}\\
& C_{f,l}\geq C_{l}^{\rm{TH}},\forall f,\forall l,\label{eq:4_5}\\
& \sum_{l=1}^{L}C_{f,l}\leq C_{f},\forall f,\label{eq:4_6}\\
& x_{f,l,k}\in\{0,1\},\forall f,\forall l,\forall k.\label{eq:4_7}
\end{align}
\end{subequations}

\noindent Constraint (\ref{eq:4_2}) indicates that one content layer is supposed to
be cached in no more than one cluster of the SBSs.
The cache size constraint of each SBS is captured in (\ref{eq:4_3}),
and the total cache size of the entire network is limited by (\ref{eq:4_4}).
To guarantee the required video contents can be correctly decoded,
the size of each content layer should be no less than the predefined level.
Then, we have constraint (\ref{eq:4_5}),
where $C_{l}^{\rm{TH}}$ denotes the minimum size requirement of the $l$-th layer.
In constraint (\ref{eq:4_6}), we ensure that
the total size of layers from the same video content will not exceed the predefined content size.
Finally, constraint (\ref{eq:4_7}) gives the feasible variable range of $x_{f,l,k}$.

Problem (\ref{max_4}) is a mixed integer programming problem,
which is NP-hard and even less tractable than Problem (\ref{max_2}).
To solve this problem, we divide (\ref{max_4}) into two subproblems,
namely, the layer placement subproblem, and the cache size and layer size allocation subproblem.
First, under the given cache size and layer size, the layer placement subproblem can be written as
\begin{subequations}\label{max_5}
\begin{align}
\underset{\mathbf{x}}{\mathrm{max}}\quad & \mathrm{ECE}\label{eq:5_1}\\
\mathrm{\mathrm{s.t.}}\:\quad & (\ref{eq:4_2}), (\ref{eq:4_3}), (\ref{eq:4_7}).
\end{align}
\end{subequations}
\noindent Obviously, Problem (\ref{max_5}) is of resemblance to Problem (\ref{max_2}).
With the help of Algorithm 1, Problem (\ref{max_5}) can be efficiently solved, where $m_{k}=M_{k}/S_{k}$ in Step 1).

Then, when the layer caching indicators are known,
the cache size and layer size allocation subproblem is given by
\begin{subequations}\label{max_6}
\begin{align}
\underset{\mathbf{M},\mathbf{C}}{\mathrm{max}}\quad & \mathrm{ECE}\label{eq:6_1}\\
\mathrm{\mathrm{s.t.}}\:\quad & (\ref{eq:4_3}), (\ref{eq:4_4}), (\ref{eq:4_5}), (\ref{eq:4_6}).
\end{align}
\end{subequations}
\noindent Problem (\ref{max_6}) is convex,
and can be efficiently solved by running the CVX solver, such as SeDuMi.
The two subproblems require to be iteratively solved until convergence.
The detailed procedures for solving Problem (\ref{max_4}) is summarized in Algorithm 2.

\section{Simulation Results}
\begin{figure*}[t]
\centering{}\subfloat[$\Pr(\mathrm{SIR}_{1}\geq\gamma_{\rm{s}})$ versus varying $\gamma_{\rm{s}}$ under
 different $S_{1}$.]{\includegraphics[scale=0.57]{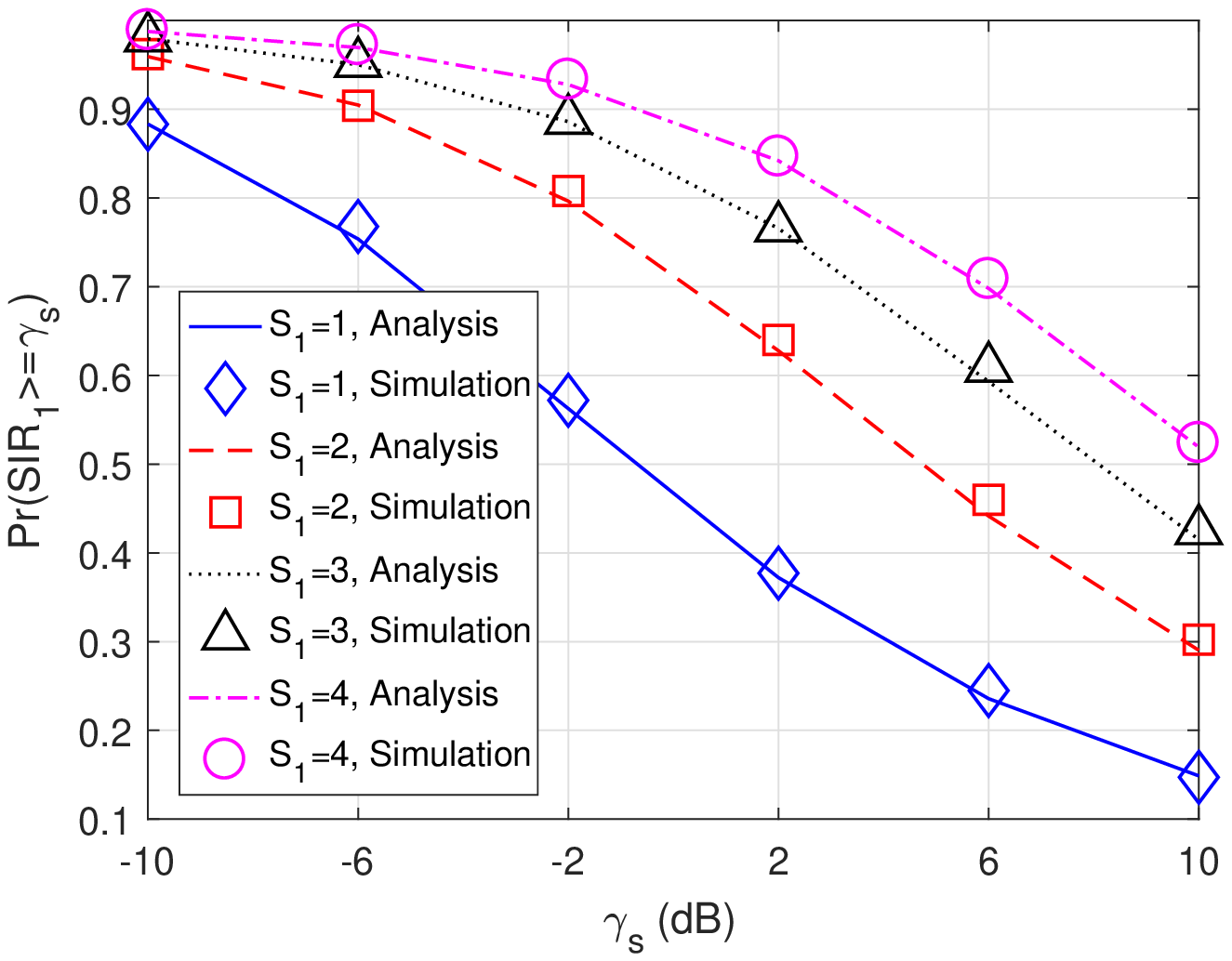}

}\subfloat[$\Pr(\mathrm{SIR}_{m_{0}}\geq\gamma_{\rm{m}})$ versus varying $\gamma_{\rm{m}}$
   under different $\alpha_{\rm{m}}$.]{\includegraphics[scale=0.57]{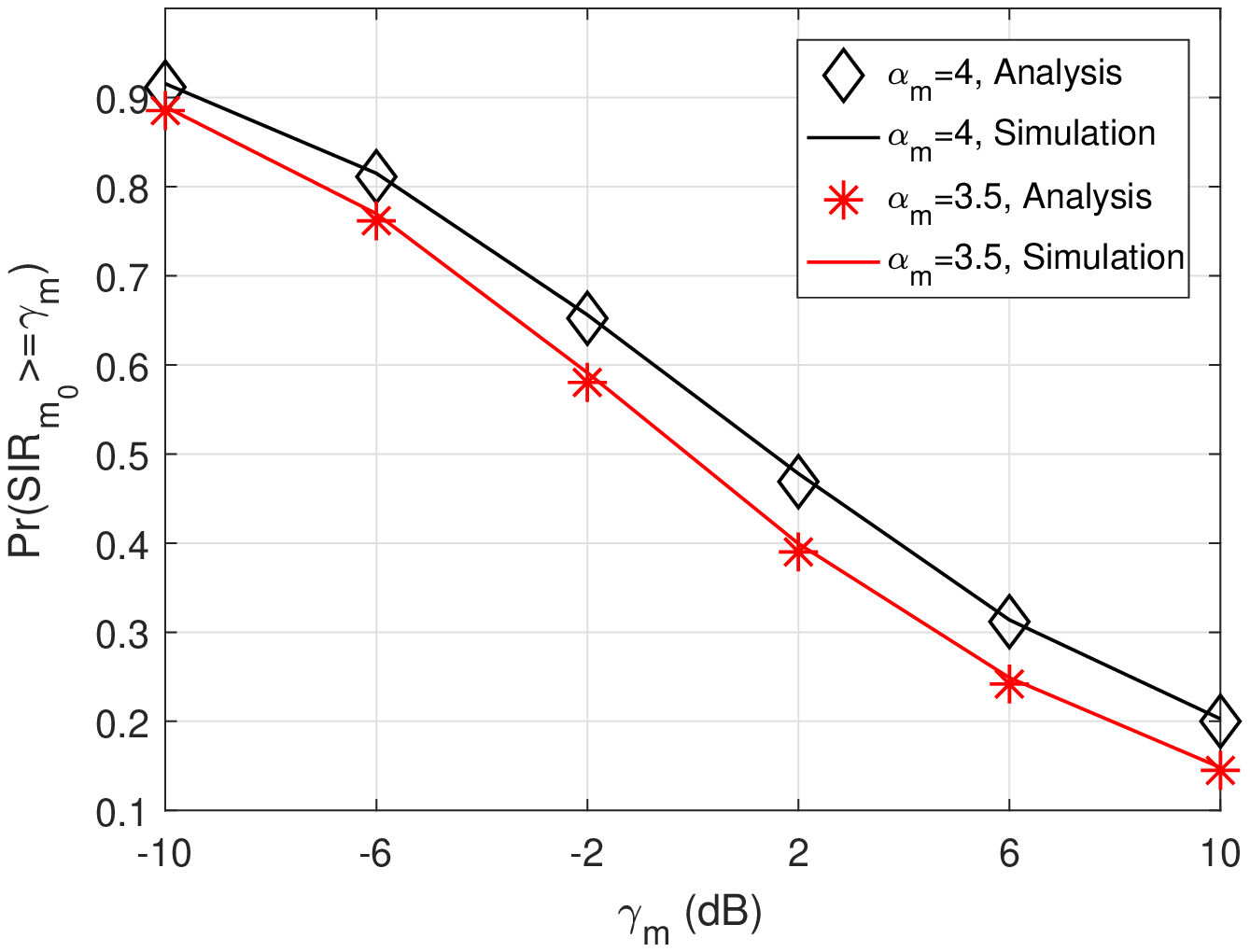}
}

\caption{The STPs when the user is served by cooperative SBSs and the nearest MBS.}
\label{Successful_Transmission_Probability}
\end{figure*}
In this section,
the simulation hardware platform is a Windows x64 system with 3.2 GHz CPU and 8 GB RAM.
Additionally, the software platform is MATLAB R2016a with off-the-shelf CVX solver, namely, SeDuMi.
We first show the simulation results of derived STPs and ESRs.
Next, the ECE performance of Algorithms 1 and 2 is demonstrated under different parameter settings.
In order to present the effectiveness of devised SVC-based caching policies,
we employ the most popular layer placement (MPLP) scheme as the comparison scheme.
In the MPLP strategy, content layers from the most popular video files are selected and locally cached in SBSs
until the cache sizes of SBSs are fully occupied.
The MPLP is able to fully utilize the local caching resources of BSs,
and is widely used in state-of-the-art researches\cite{Tao2015Content, Liu2016Energy}.
Additionally, for very few studies which jointly considered SVC with wireless caching \cite{ZhanContent, Xiang2018Cache},
the cache placement policies were tailored for their specific scenarios,
which are substantially different from ours.
They cannot provide meaningful comparisons to our approaches.
As a result, we focus on MPLP as the benchmark.
The settings of simulation parameters are listed in Table I.

\begin{table}[t]
\centering{}

\caption{Values of Simulation Parameters}

\begin{tabular}{c|c}
\hline
Parameter & Value\tabularnewline
\hline
$K$& $3$\tabularnewline
\hline
$P_{\rm{s}}$, $P_{\rm{m}}$ & $23$ dBm, $33$ dBm\tabularnewline
\hline
$\lambda_{\rm{s}},\lambda_{\rm{m}}$ & $1/(100^{2}\pi) /\rm{m}^{2}$, $1/(500^{2}\pi) /\rm{m}^{2}$\tabularnewline
\hline
$d_{1}$, $d_{2}$, $d_{3}$& 10 m, 20 m, 50 m\tabularnewline
\hline
$\alpha_{\rm{s}}$, $\alpha_{\rm{m}}$ & $4$\tabularnewline
\hline
$M$ & $1000$ Mbits \tabularnewline
\hline
$C_{f}$ & $5
0$ Mbits \tabularnewline
\hline
$F$ & 100\tabularnewline
\hline
$L$ & 5\tabularnewline
\hline
$\alpha$ & $1$ \tabularnewline
\hline
$W_{\rm{s}}$, $W_{\rm{m}}$ & $10$ MHz, $50$ MHz \tabularnewline
\hline
$\gamma_{\rm{s}}$, $\gamma_{\rm{m}}$ & $10$ dB, $5$ dB \tabularnewline
\hline
$P_{\rm{fix,s}},P_{\rm{fix,m}}$ & $6.8$ W, $30 $ W\tabularnewline
\hline
$k_{\rm{c}}$   & $3.87\times10^{-4}$ CNY/Joule \tabularnewline
\hline
$k_{\rm{r}}$   & $1.41\times10^{-8}$ CNY/bit \tabularnewline
\hline
$\tau$   & $10^{-11}$  \tabularnewline
\hline
\end{tabular}
\end{table}

Fig. \ref{Successful_Transmission_Probability} illustrates the numerical results of derived expressions,
as well as the Monte Carlo simulations, for STPs under varying QoS requirements.
From the figure, we can see that the gap between theoretical analysis and Monte Carlo simulations is negligible,
which validates our derived expressions presented in Theorem 2.
Additionally, it is obvious that higher QoS requirement results in poorer STP performance,
while increasing the number of serving SBSs in a cluster will have a positive impact on STP.
From Fig. 2 (b), we can conclude that lower pathloss exponent cannot improve the STP,
since the increase of received signal strength cannot make up for the loss caused by the increased interference.

In Fig. \ref{Ergodic_Service_Rate}, we present the numerical results of derived expressions for ESR
derived in (\ref{SBS_Rate}) and (\ref{MBS_Rate}) under varying QoS requirements.
As $\gamma_{\rm{s}}$ and $\gamma_{\rm{m}}$ grow,
though the STP decreases, the ESR increases.
Similar to Fig. 2, more serving SBSs in a cluster will lead to larger ESR,
since the STP increases as the number of serving SBSs grows.

\begin{figure*}[t]
\centering{}\subfloat[$R_{k}$ versus varying $\gamma_{\rm{s}}$ under
 different $k$ and $S_{k}$.]{\includegraphics[scale=0.57]{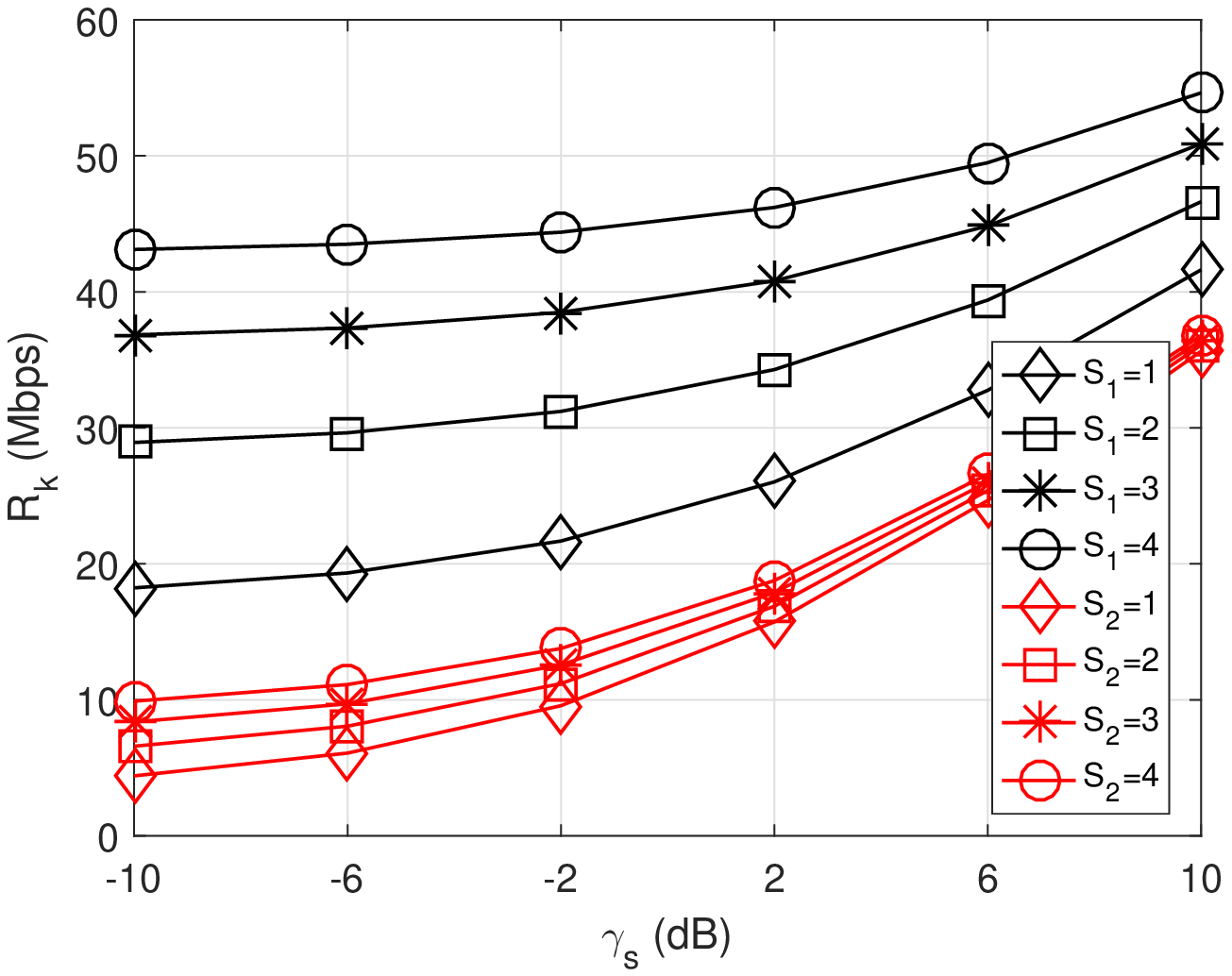}

}\subfloat[$R_{m_{0}}$ versus varying $\gamma_{\rm{m}}$
   under different $\alpha_{\rm{m}}$.]{\includegraphics[scale=0.57]{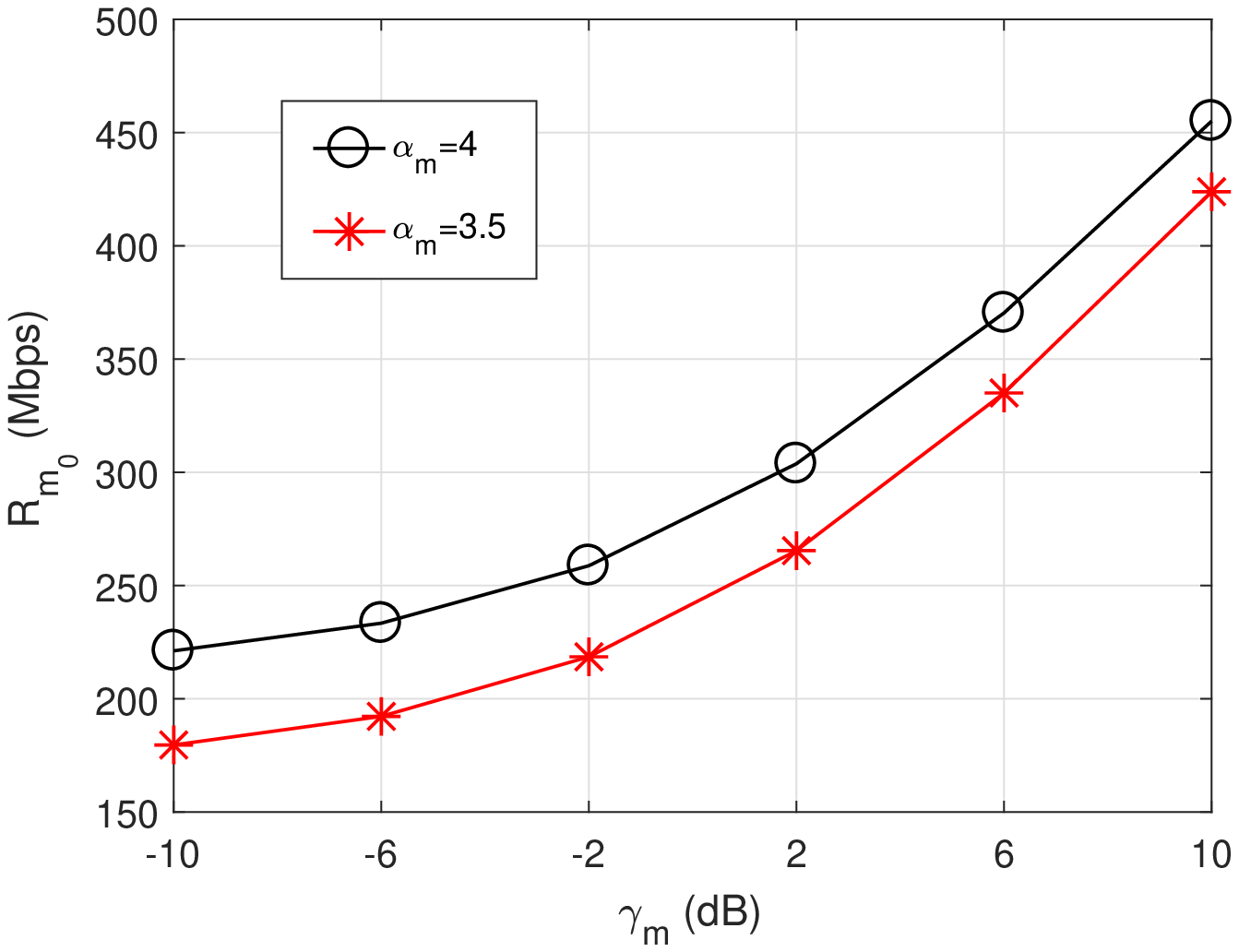}
}

\caption{The ESRs when the typical user is served by cooperative SBSs and the nearest MBS.}
\label{Ergodic_Service_Rate}
\end{figure*}

In Fig. \ref{ECE_Gamma}, we present the relationship between the ECE performance and the minimum QoS requirement $\gamma_{\rm{s}}$.
It is apparent that a larger value of $\gamma_{\rm{s}}$ leads to a better ECE performance,
since the ESR increases as $\gamma_{\rm{s}}$ grows.
It is noted that the optimal ECE derived from the relaxed Problem (\ref{max_3}) can be regarded as the upper bound of Algorithm 1,
since the feasible range of optimization variables becomes larger.
From this figure, we also find that the ECE of Algorithm 1 is almost the same as the upper bound obtained from Problem (\ref{max_3}),
which demonstrates the optimality of Algorithm 1.
Obviously, the ECE obtained from Algorithm 2 outperforms the ECE derived from Algorithm 1,
since the cache size and layer size can be adaptively adjusted in response to different user requirements and network environments.
Both of the proposed algorithms show better ECE performance than MPLP scheme.
The reason for this is given as follows.
The proposed SVC-based layer caching schemes fully exploit the layer diversity,
such that more required layers can be found in the local cache of SBSs,
avoiding repeated backhaul deliveries and reducing extra resource cost.

\begin{figure}[t]
\begin{center}
\centering{}\includegraphics[scale=0.6]{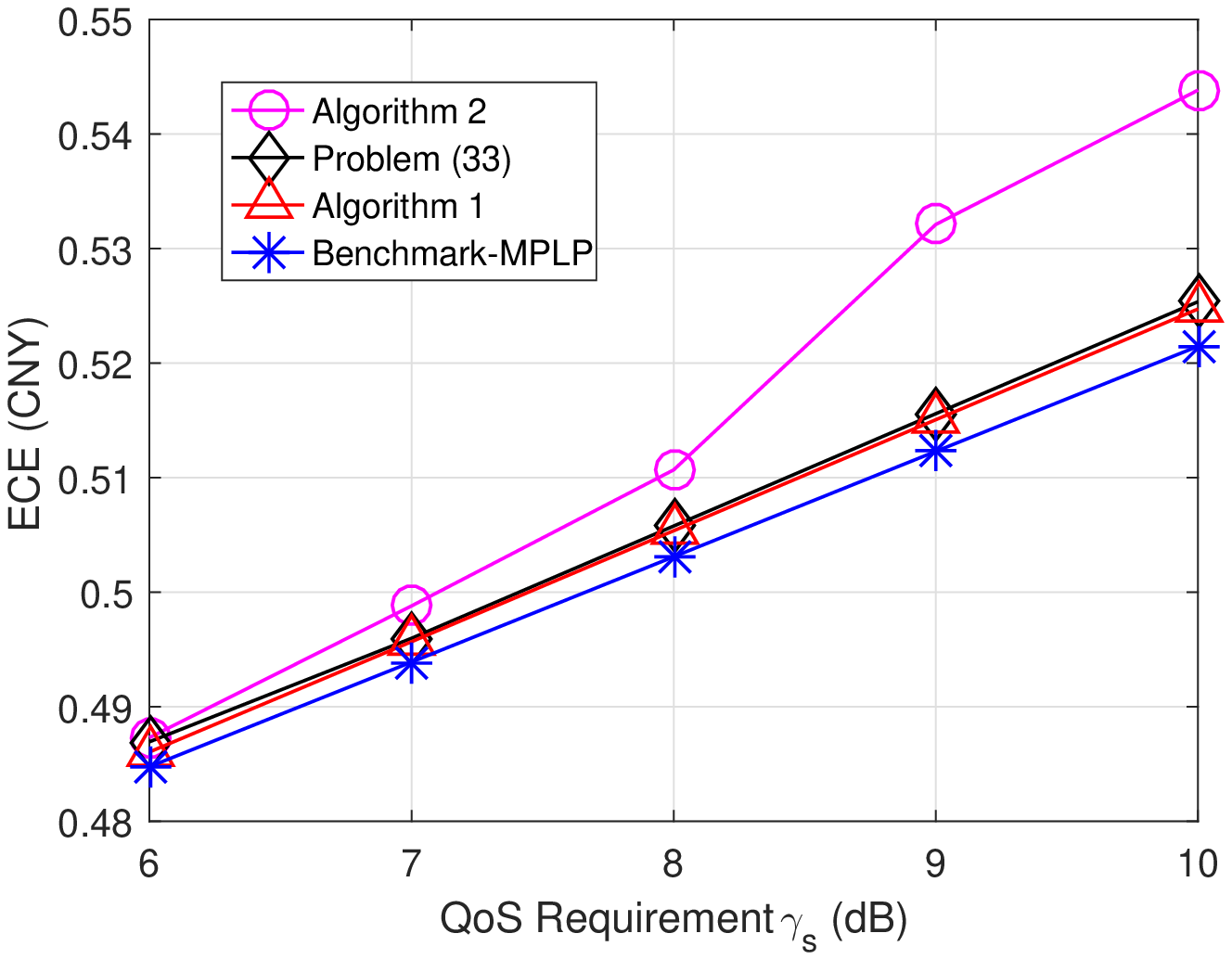}
\caption{The ECE performance versus the minimum QoS requirement $\gamma_{\rm{s}}$.}
\label{ECE_Gamma}
\end{center}
\end{figure}

From Fig. \ref{ECE_Skewness}, we can find the relationship between the ECE performance and the skewness parameter $\alpha$.
As described earlier, a larger value of $\alpha$ typically leads to the fact
that fewer video contents can satisfy the majority of multimedia video requests.
As $\alpha$ increases, the performance gaps between the proposed algorithms and MPLP diminish.
This is because that the user requests are increasingly concentrated on the top popular videos,
and a few of them can satisfy most of the user requests.
When the skewness parameter $\alpha$ is large enough,
performance gaps between the proposed algorithms and MPLP will disappear,
and the proposed caching schemes are equivalent to the MPLP strategy.

\begin{figure}[t]
\centering{}\includegraphics[scale=0.6]{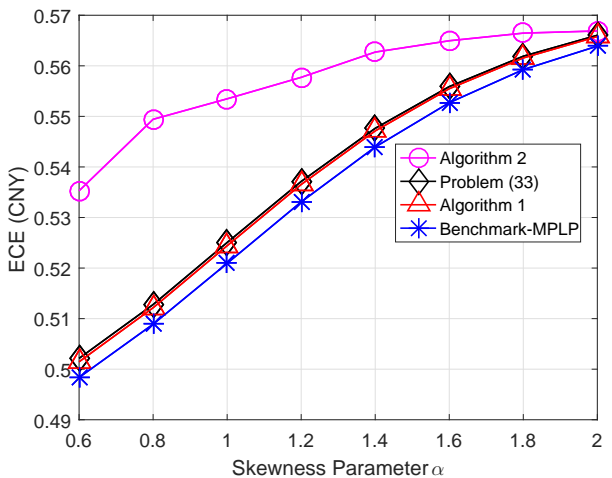}
\caption{The ECE performance versus the skewness parameter $\alpha$.}
\label{ECE_Skewness}
\end{figure}

In Fig. \ref{ECE_Cachesize}, we present the relationship between the ECE performance and the cache size $M$.
With larger cache size equipped at each SBS, more video layers can be locally cached.
Additionally, repeated deliveries through backhaul links can be avoided,
thus the ECE performance is improved.
On the other hand, with more layers placed in the local cache of SBSs,
content layers can be immediately transmitted to the user if requested,
reducing the service delay and thus enhancing the quality of experience (QoE) of mobile users.

\begin{figure}[t]
\centering{}\includegraphics[scale=0.6]{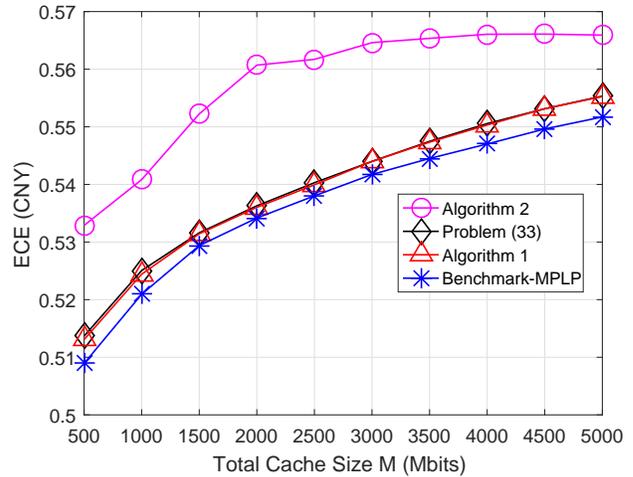}
\caption{The ECE performance versus the total cache size $M$.}
\label{ECE_Cachesize}
\end{figure}

In Fig. \ref{ECE_Backhaul},
the relationship between the ECE performance and the backhaul power coefficient $c_{\rm{bh}}$ is presented.
It is apparent that a larger value of $c_{\rm{bh}}$ has a more negative impact on ECE.
As $c_{\rm{bh}}$ grows, the backhaul delivery will be more power-consuming,
thus increasing the total cost and degrading the ECE performance.
In the power-limited case, the wired backhaul can be employed, though the deployment flexibility is poor.
Otherwise, the wireless backhaul, i.e., the optical backhaul, can be applied.

\begin{figure}[t]
\centering{}\includegraphics[scale=0.6]{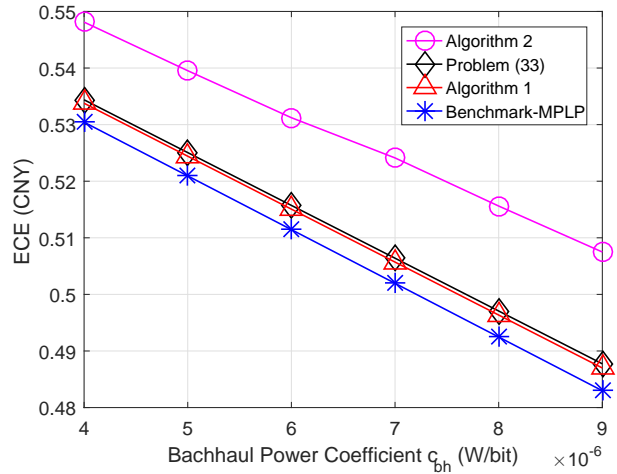}
\caption{The ECE performance versus the backhaul power coefficient $c_{\rm{bh}}$.}
\label{ECE_Backhaul}
\end{figure}

Fig. \ref{ECE_LayerNumber} presents the relationship between the ECE performance and the number of divided content layers $L$.
As we can see, the ECE increases as $L$ grows,
since more video layers can be adaptively placed and the limited cache size of SBSs can be fully exploited.
Notably, when $L$ becomes large,
the performance gap between Algorithm 1 and the relaxed Problem (\ref{max_3}) diminishes,
which validates the optimality of Algorithm 1 given in Theorem 4.
In the practical scenario, when the value of $FLK$ is considerably large,
the optimal layer caching indicators resulting from (\ref{max_3}) can be optimal.

\begin{figure}[t]
\centering{}\includegraphics[scale=0.57]{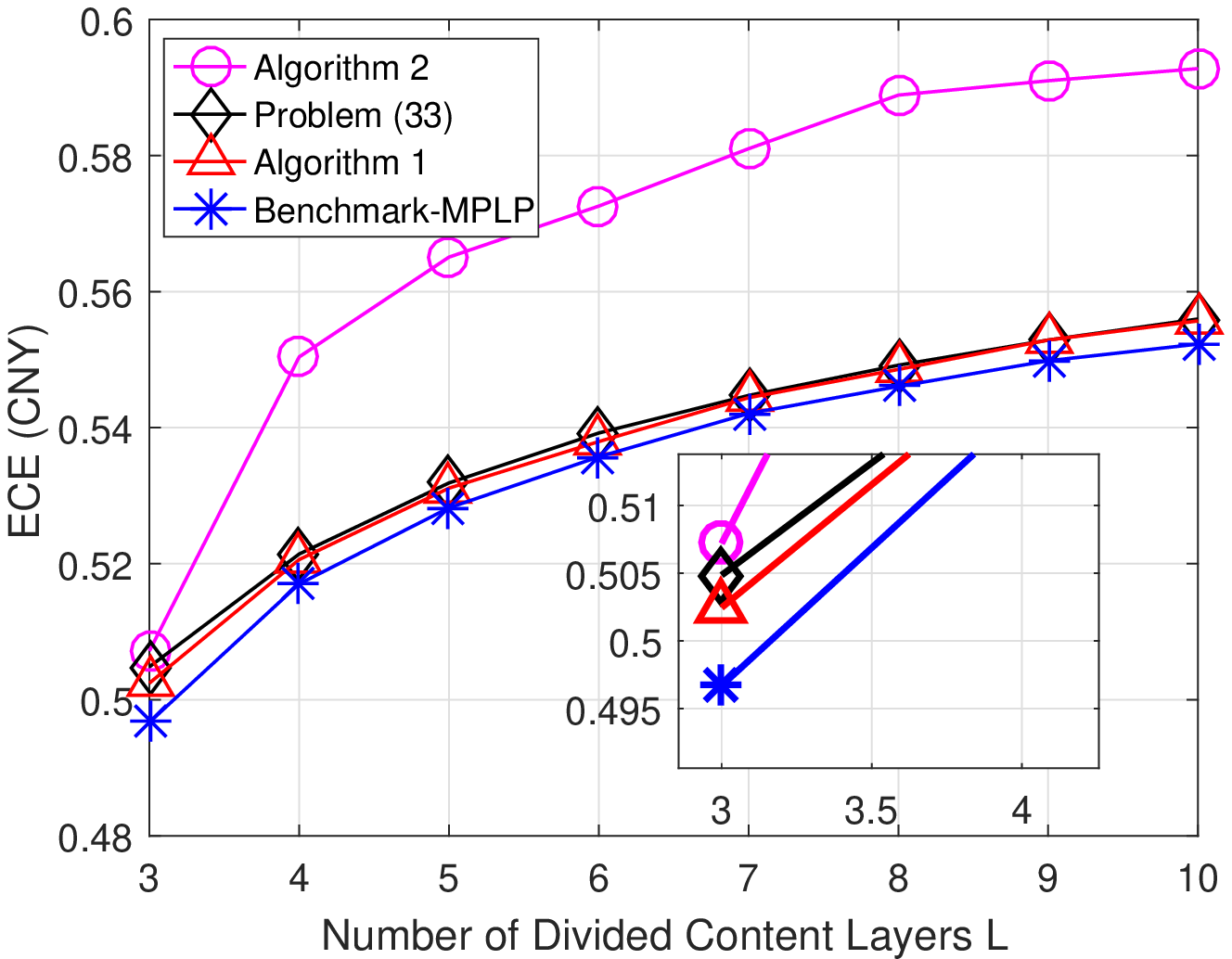}
\caption{The ECE performance versus the number of divided content layers $L$.}
\label{ECE_LayerNumber}
\end{figure}
\section{Conclusion}
In this paper, we investigate the optimal economical caching schemes
in cache-enabled heterogeneous networks to provide SDV and HDV services to mobile users.
In order to enhance the ECE performance,
we formulate the ECE optimization problems for two cases.
In the first case, equal cache size is equipped at each SBS.
The original problem is first relaxed to be convex after the $l_{0}$-norm approximation.
Then, a greedy-strategy based algorithm is proposed to achieve the near-optimal layer caching indicators.
In the second case, with unequal cache size and layer size,
the ECE maximization problem is divided into two subproblems,
which are then independently solved.
Simulation results validate the theoretical analysis,
and confirm the effectiveness of the proposed economical caching schemes as compared to the MPLP strategy.
Moreover, the optimality of the proposed greedy-strategy based algorithm is confirmed.
\begin{appendices}
\section{Proof of Theorem 2}
First, the expressions for STPs can be re-written as
\begin{align}
\Pr(\mathrm{SIR}_{k}\geq\gamma_{\rm{s}})=&\int_{d_{k-1}}^{d_{k}}...\int_{d_{k-1}}^{d_{k}}f_{\rm{s}}(\mathbf{x}_{k})\nonumber\\
&\Pr(\mathrm{SIR}_{k}\geq\gamma_{\rm{s}}|\mathbf{r}_{k}=\mathbf{x}_{k})\mathrm{d}\mathbf{x}_{k}\label{asd},
\end{align}
\begin{align}
\Pr(\mathrm{SIR}_{m_{0}}\geq\gamma_{\rm{m}})=&\int_{0}^{\infty}f_{\rm{m}}(x_{m_{0}})\nonumber\\
&\Pr(\mathrm{SIR}_{m_{0}}\geq\gamma_{\rm{m}}|r_{m_{0}}=x_{m_{0}})\mathrm{d}x_{m_{0}}\label{dfg}.
\end{align}
For simplicity of notations, we define
\begin{align}
&I_{\rm{s}}=\sum_{s\in\mathrm{\Phi_{\rm{s}}}\mathrm{\setminus}\mathcal{S}_{1},...,\mathcal{S}_{k}}\left|h_{s}\right|^{2}r_{s}^{-\alpha_{\rm{s}}},\\
&I_{\rm{m}}=\sum_{m\in\mathrm{\Phi_{\rm{m}}\setminus}m_{0}}\left|h_{m}\right|^{2}r_{m}^{-\alpha_{\rm{m}}}.
\end{align}

\noindent The conditional STP for SBSs in cluster $\mathcal{S}_{k}$ can be calculated as
\begin{align}
 & \Pr(\mathrm{SIR}_{k}\geq\gamma_{\rm{s}}|\mathbf{r}_{k}=\mathbf{x}_{k})\nonumber\\
 & =\Pr(\frac{\sum_{l=S_{k-1}+1}^{S_{k}}\left|h_{l}x_{l}^{-\frac{\alpha_{\rm{s}}}{2}}\right|^{2}}{\sum_{s\in\mathrm{\Phi_{\rm{s}}}\mathrm{\setminus}\mathcal{S}_{1},...,\mathcal{S}_{k}}\left|h_{s}\right|^{2}r_{s}^{-\alpha_{\rm{s}}}}\geq\gamma_{\rm{s}})\nonumber
 \end{align}
 \begin{align}
 &=\Pr(\sum_{l=S_{k-1}+1}^{S_{k}}\left|h_{l}x_{l}^{-\frac{\alpha_{\rm{s}}}{2}}\right|^{2}\geq\gamma_{\rm{s}}I_{\rm{s}})\nonumber\\
 & \stackrel{(a)}{=}\mathbb{E}_{I_{\rm{s}}}[\exp(-\frac{1}{\sum_{l=S_{k-1}+1}^{S_{k}}x_{l}^{-\alpha_{\rm{s}}}}\gamma_{\rm{s}}I_{\rm{s}})]\nonumber\\
 & =\mathcal{L}_{I_{\rm{s}}}(\frac{\gamma_{\rm{s}}}{\sum_{l=S_{k-1}+1}^{S_{k}}x_{l}^{-\alpha_{\rm{s}}}}).
\end{align}

\noindent Equality (a) holds since $\sum_{l=S_{k-1}+1}^{S_{k}}\left|h_{l}x_{l}^{-\frac{\alpha_{\rm{s}}}{2}}\right|^{2}\sim\exp(\frac{1}{\sum_{l=S_{k-1}+1}^{S_{k}}x_{l}^{-\alpha_{\rm{s}}}})$,
where $\exp(\lambda)$ denotes the exponential distribution with mean $\lambda$.
For notational convenience, we denote $k_{1}=\frac{1}{\sum_{l=S_{k-1}+1}^{S_{k}}x_{l}^{-\alpha_{\rm{s}}}}$.
Besides, $\mathcal{L}_{I_{\rm{s}}}(\frac{\gamma_{\rm{s}}}{\sum_{l=S_{k-1}+1}^{S_{k}}x_{l}^{-\alpha_{\rm{s}}}})$
is the Laplace transform of the interference generated by the PPP-distributed SBSs that are farther than those in $\mathcal{S}_{k}$,
which is calculated as
\begin{align}
 & \mathcal{L}_{I_{\rm{s}}}(\frac{\gamma_{\rm{s}}}{\sum_{l=S_{k-1}+1}^{S_{k}}x_{l}^{-\alpha_{\rm{s}}}})\nonumber \\
 & =\mathbb{E}_{\Phi_{\rm{s}},h_{s}}[\prod_{s\in\mathrm{\Phi_{\rm{s}}}\mathrm{\setminus}\mathcal{S}_{1},...,\mathcal{S}_{k}}\exp(-k_{1}\gamma_{\rm{s}}\left|h_{s}\right|^{2}r_{s}^{-\alpha_{\rm{s}}})]\nonumber\\
 & =\mathbb{E}_{\Phi_{\rm{s}}}[\prod_{s\in\mathrm{\Phi_{\rm{s}}}\mathrm{\setminus}\mathcal{S}_{1},...,\mathcal{S}_{k}}\frac{1}{k_{1}\gamma_{\rm{s}}r_{s}^{-\alpha_{\rm{s}}}}]\nonumber\\
 & \stackrel{(b)}{=}\exp(-2\pi\lambda_{\rm{s}}\int_{d_{k}}^{\infty}(1-\frac{1}{1+k_{1}\gamma_{\rm{s}}r_{s}^{-\alpha_{\rm{s}}}})r_{s}\mathrm{d}r_{s})\nonumber \\
 & =\exp(-\pi\lambda_{\rm{s}}(k_{1}\gamma_{\rm{s}})^{\frac{2}{\alpha_{\rm{s}}}}G_{\alpha_{\rm{s}}}(d_{k}^{2}(k_{1}\gamma_{\rm{s}})^{-\frac{2}{\alpha_{\rm{s}}}})),\label{eq:aaa}
\end{align}
\noindent where $G_{a}(b)=\int_{b}^{\infty}\frac{1}{1+r^{\frac{a}{2}}}\mathrm{d}r$
and equality (b) holds due to the property of the probability generating functional (PGF) of the PPP process ${\rm{\Phi_{\rm{s}}}}$.
Substituting (\ref{eq:aaa}) into (\ref{asd}),
we can obtain the STP $\Pr(\mathrm{SIR}_{k}\geq\gamma_{\rm{s}})$.

Next, the conditional probability $\Pr(\mathrm{SIR}_{m_{0}}\geq\gamma_{\rm{m}}|r_{m_{0}}=x_{m_{0}})$ can be calculated as
\begin{align}
 & \Pr(\mathrm{SIR}_{m_{0}}\geq\gamma_{\rm{m}}|r_{m_{0}}=x_{m_{0}})\nonumber\\
 & =\Pr(\frac{\left|h_{m_{0}}\right|^{2}x_{m_{0}}^{-\alpha_{\rm{m}}}}{\sum_{m\in\mathrm{\Phi_{\rm{m}}\setminus}m_{0}}\left|h_{m}\right|^{2}r_{m}^{-\alpha_{\rm{m}}}}\geq\gamma_{\rm{m}})\nonumber
\end{align}
\begin{align}
 & =\Pr(\left|h_{m_{0}}\right|^{2}\geq\gamma_{\rm{m}}x_{m_{0}}^{\alpha_{\rm{m}}}I_{\rm{m}})\nonumber\\
 & \stackrel{(c)}{=}\mathbb{E}_{I_{\rm{m}}}[\exp(-\gamma_{\rm{m}}x_{m_{0}}^{\alpha_{\rm{m}}}I_{\rm{m}})]\nonumber\\
 & =\mathcal{L}_{I_{\rm{m}}}(\gamma_{\rm{m}}x_{m_{0}}^{\alpha_{\rm{m}}}),
\end{align}
\noindent where (c) holds because $\left|h_{m_{0}}\right|^{2}\sim\exp(1)$.
Following the similar steps, $\mathcal{L}_{I_{\rm{m}}}(\gamma_{\rm{m}}x_{m_{0}}^{\alpha_{\rm{m}}})$ can be derived as
\begin{align}
 &\mathcal{L}_{I_{\rm{m}}}(\gamma_{\rm{m}}x_{m_{0}}^{\alpha_{\rm{m}}})\nonumber \\
 &=\mathbb{E}_{\Phi_{\rm{m}},h_{m}}[\prod_{m\in\mathrm{\Phi_{\rm{m}}\setminus}m_{0}}\exp(-\gamma_{\rm{m}}x_{m_{0}}^{\alpha_{\rm{m}}}\left|h_{m}\right|^{2}r_{m}^{-\alpha_{\rm{m}}})]\nonumber\\
 & =\mathbb{E}_{\Phi_{\rm{m}}}[\prod_{m\in\mathrm{\Phi_{\rm{m}}\setminus}m_{0}}\frac{1}{\gamma_{\rm{m}}x_{m_{0}}^{\alpha_{\rm{m}}}r_{m}^{-\alpha_{\rm{m}}}}]\nonumber\\
 & =\exp(-2\pi\lambda_{\rm{m}}\int_{x_{m_{0}}}^{\infty}(1-\frac{1}{1+\gamma_{\rm{m}}x_{m_{0}}^{\alpha_{\rm{m}}}r_{m}^{-\alpha_{\rm{m}}}})r_{m}\mathrm{d}r_{m})\nonumber \\
 & =\exp(-\pi\lambda_{\rm{m}}\gamma_{\rm{m}}^{\frac{2}{\alpha_{\rm{m}}}}x_{m_{0}}^{2}G_{\alpha_{\rm{m}}}(\gamma_{\rm{m}}^{-\frac{2}{\alpha_{\rm{m}}}})).\label{eq:bbb}
\end{align}
\noindent Then, $\Pr(\mathrm{SIR}_{m_{0}}\geq\gamma_{\rm{m}})$ can be derived as
\begin{align}
&\Pr(\mathrm{SIR}_{m_{0}}\geq\gamma_{\rm{m}})=2\pi\lambda_{m}\nonumber\\
&\int_{0}^{\infty}x_{m_{0}}\exp(-\pi\lambda_{\rm{m}}x_{m_{0}}^{2}(\gamma_{\rm{m}}^{\frac{2}{\alpha_{\rm{m}}}}G_{\alpha_{\rm{m}}}(\gamma_{\rm{m}}^{-\frac{2}{\alpha_{\rm{m}}}})+1))\mathrm{d}x_{m_{0}}\nonumber\\
&=\pi\lambda_{m}\int_{0}^{\infty}\exp(-\pi\lambda_{\rm{m}}x_{m_{0}}^{2}(\gamma_{\rm{m}}^{\frac{2}{\alpha_{\rm{m}}}}G_{\alpha_{\rm{m}}}(\gamma_{\rm{m}}^{-\frac{2}{\alpha_{\rm{m}}}})+1))\mathrm{d}x_{m_{0}}^{2}\nonumber\\
&=(\gamma_{\rm{m}}^{\frac{2}{\alpha_{\rm{m}}}}G_{\alpha_{\rm{m}}}(\gamma_{\rm{m}}^{-\frac{2}{\alpha_{\rm{m}}}})+1)^{-1}.
\end{align}
\qed
\section{Proof of Theorem 3}
In Algorithm 1, from Steps 6) to 11),
we find that each content layer is cached in the SBS that can provide the largest marginal ECE,
evaluated by $z_{f,l,k}$, if the reserved storage space is large enough.
This means that, in the case of limited cache size at each SBS,
the devised layer-SBS associations, i.e., the layer placement, are the best choices at the current conditions,
and these associations will not be broken.
Therefore, it can be inferred that the proposed layer placement scheme is stable.
\qed
\end{appendices}

\bibliographystyle{IEEEtran}

\bibliography{ciations}

\begin{IEEEbiography}[{\includegraphics[width=1in,height=1.25in,clip,keepaspectratio]{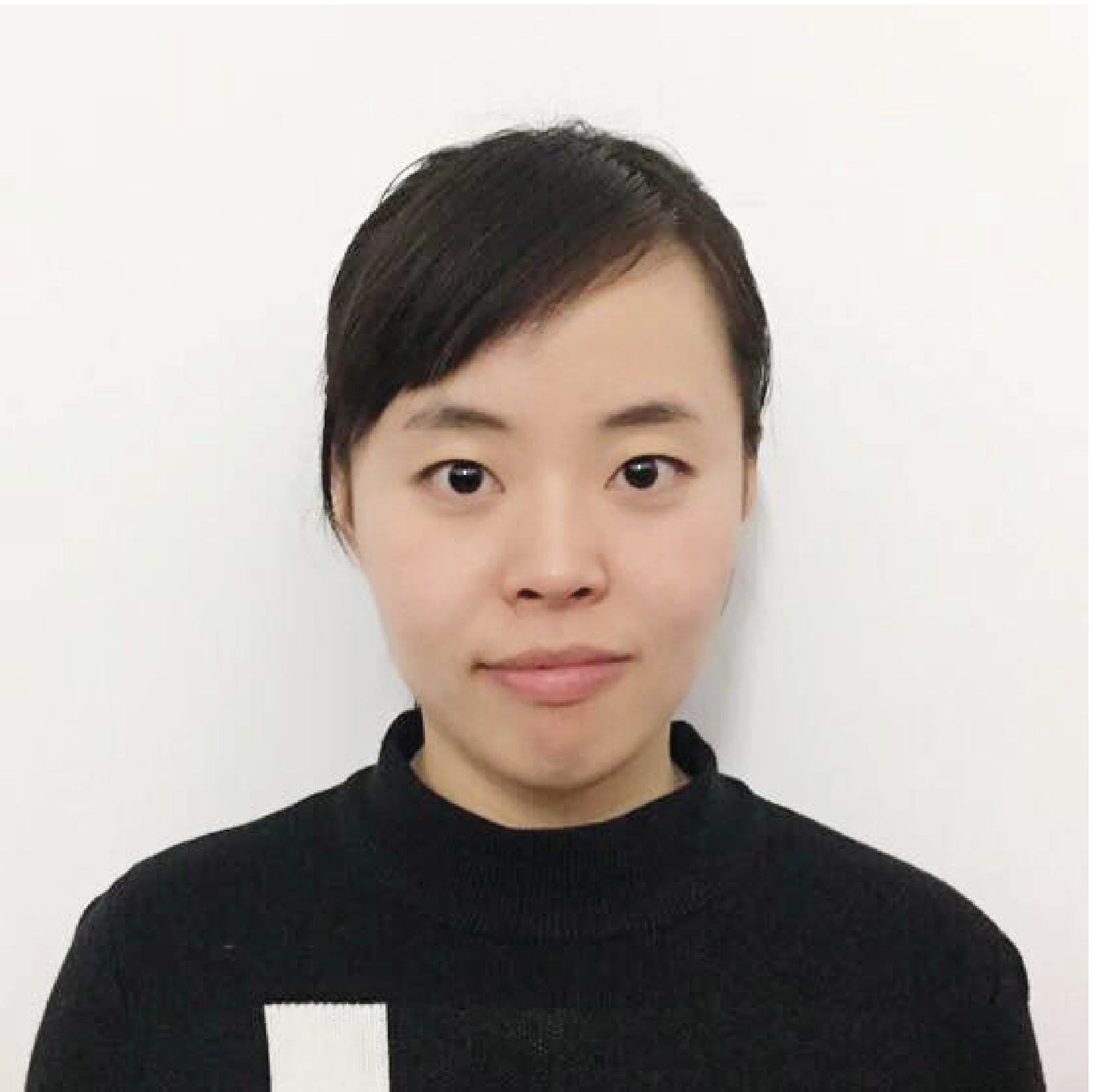}}]{Xuewei Zhang}

(S'18) received the B.E. degree in communication engineering from Tianjin Polytechnic University, Tianjin, China, in 2015.
She is currently pursuing the Ph.D. degree with the School of Information and Communication Engineering,
Beijing University of Posts and Telecommunications (BUPT), Beijing, China.
Her research interests include multicast beamforming, wireless caching and resource allocation in heterogeneous networks.

\end{IEEEbiography}

\begin{IEEEbiography}[{\includegraphics[width=1in,height=1.25in,clip,keepaspectratio]{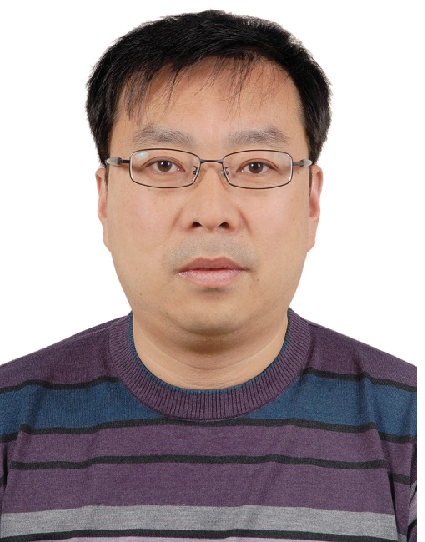}}]{Tiejun Lv}

(M'08-SM'12) received the M.S. and Ph.D. degrees in electronic engineering from
the University of Electronic Science and Technology of China (UESTC), Chengdu, China, in 1997 and 2000, respectively.
From January 2001 to January 2003, he was a Postdoctoral Fellow with Tsinghua University, Beijing, China.
In 2005, he was promoted to a Full Professor with the School of Information and Communication Engineering,
Beijing University of Posts and Telecommunications (BUPT).
From September 2008 to March 2009, he was a Visiting Professor with the Department of Electrical Engineering, Stanford University, Stanford, CA, USA.
He is the author of 2 books, more than 60 published IEEE journal papers and 180 conference papers on the physical layer of wireless mobile communications.
His current research interests include signal processing, communications theory and networking.
He was the recipient of the Program for New Century Excellent Talents in University Award from the Ministry of Education, China, in 2006.
He received the Nature Science Award in the Ministry of Education of China for the hierarchical cooperative communication theory and technologies in 2015.

\end{IEEEbiography}

\begin{IEEEbiography}[{\includegraphics[width=1in,height=1.25in,clip,keepaspectratio]{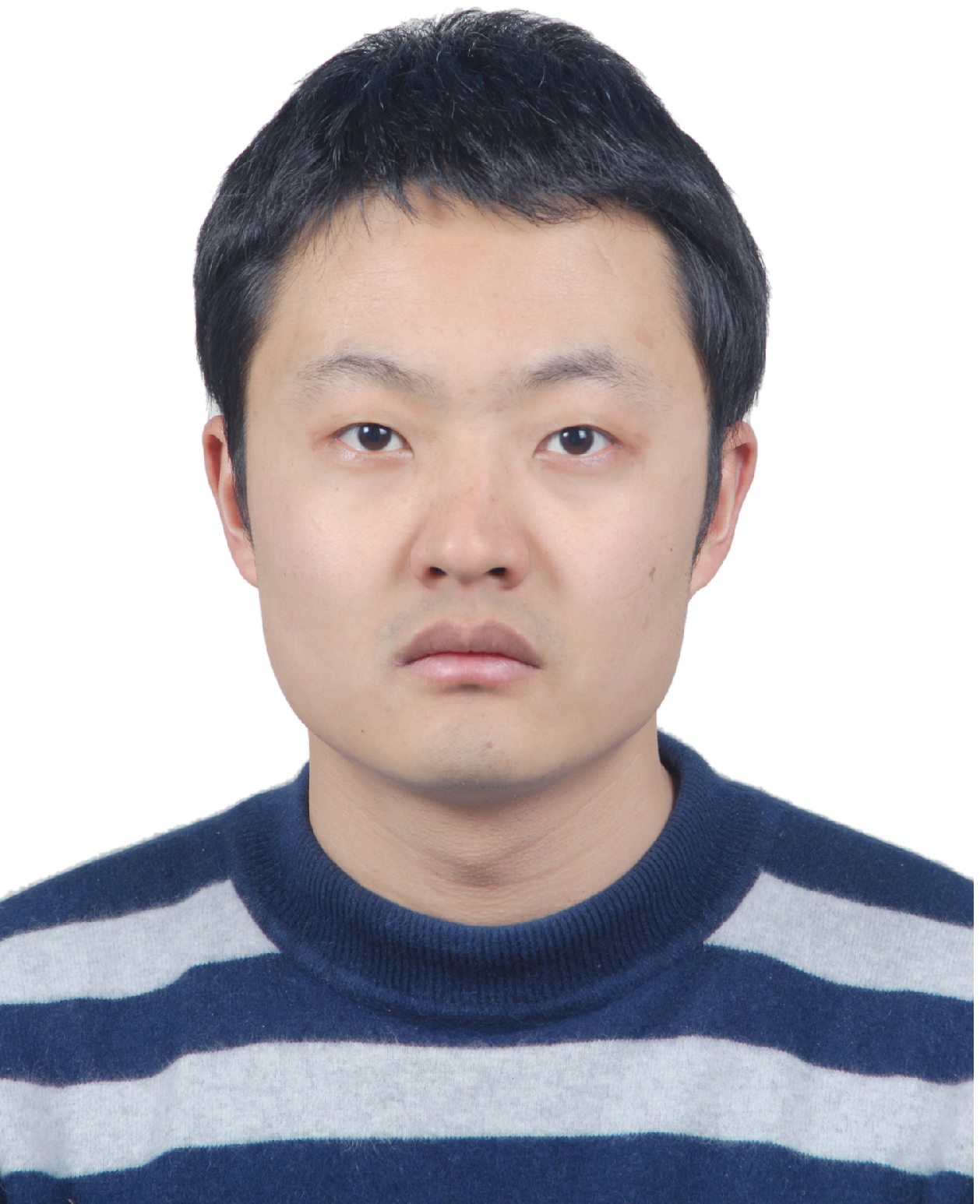}}]{Yuan Ren}

Yuan Ren received the B.Eng. degree in information engineering and the Ph.D. degree in signal and information processing
from the Beijing University of Posts and Telecommunications (BUPT), Beijing, China, in 2010 and 2017, respectively.
He is currently a Lecturer with the School of Information and Communication Engineering,
Xi'an University of Posts and Telecommunications.
His current research interests include green communications, wireless caching, and cooperative communications.

\end{IEEEbiography}

\begin{IEEEbiography}[{\includegraphics[width=1in,height=1.25in,clip,keepaspectratio]{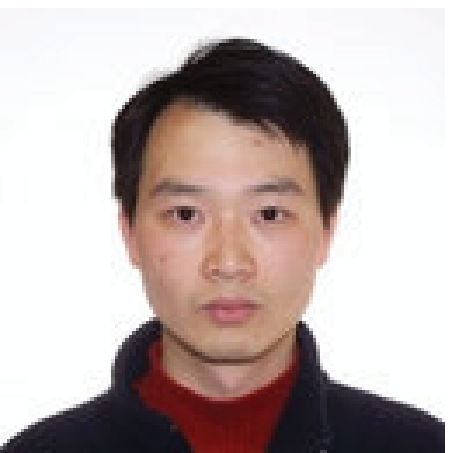}}]{Wei Ni}

 received the B.E. and Ph.D. degrees in Electronic Engineering from Fudan University, Shanghai, China, in 2000 and 2005, respectively.
 Currently he is a Team Leader, Senior Scientist, and Project Leader at CSIRO, Sydney, Australia.
 He also holds honorary positions at the University of New South Wales (UNSW), Macquarie University (MQ) and the University of Technology Sydney (UTS).
 Prior to this he was a postdoctoral research fellow at Shanghai Jiaotong University (2005 -- 2008),
 Research Scientist and Deputy Project Manager at the Bell Labs R\&I Center, Alcatel/Alcatel-Lucent (2005 -- 2008),
 and Senior Researcher at Devices R\&D, Nokia (2008-2009).
 His research interests include optimization, game theory, graph theory, as well as their applications to network and security.

Dr Ni serves as Vice Chair of IEEE NSW VTS Chapter since 2018, served as Editor for Hindawi Journal of Engineering from 2012 -- 2015, secretary of IEEE NSW VTS Chapter from 2015-2018, Track Chair for VTC-Spring 2017, Track Co-chair for IEEE VTC-Spring 2016, and Publication Chair for BodyNet 2015. He also served as Student Travel Grant Chair for WPMC 2014, a Program Committee Member of CHINACOM 2014, a TPC member of IEEE ICC14, ICCC15, EICE14, and WCNC10.

\end{IEEEbiography}

\begin{IEEEbiography}[{\includegraphics[width=1in,height=1.25in,clip,keepaspectratio]{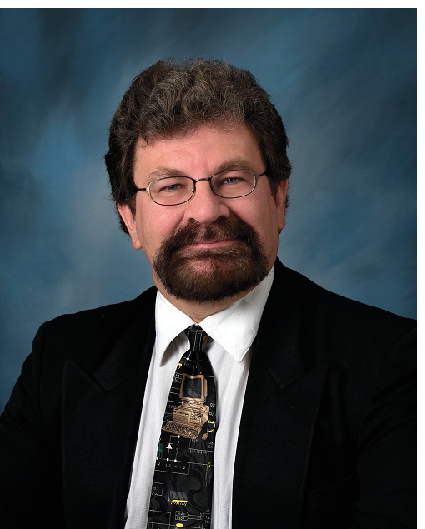}}]{Norman C. Beaulieu}

is an Academician of the Royal Society of Canada (RSC), and an Academician of the Canadian Academy of Engineering (CAE).
He is a Fellow of the Institute of Electrical and Electronics Engineers (IEEE),
a Fellow of the Institution of Engineering and Technology (IET) of the United Kingdom,
a Fellow of the Engineering Institute of Canada (EIC), and a Nicola Copernicus Fellow of Italy.
He is the recipient of the esteemed Natural Sciences and Engineering Research Council (NSERC) of Canada E.W.R. Steacy Memorial Fellowship.
He is the only person in the world to hold both the IEEE Edwin Howard Armstrong Award and the IEEE Reginald Aubrey Fessenden Award,
named for the inventors of Frequency Modulation or FM, and Amplitude Modulation or AM, respectively.
Prof. Beaulieu is a Beijing University of Posts and Telecommunications BUPT Thousand-Talents Scholar.
He holds the third highest Web of Science ISI h-index in the world in the combined areas of communication theory and information theory.
Prof. Beaulieu was awarded the title ``State Especially Recruited Foreign Expert'' certified upon him by Minister of Human Resources and Social Insurance, and Vice Minister of the Organization Department, Yi Weimin.
Pro. Beaulieu is the recipient of the Royal Society of Canada Thomas W.  Eadie Medal, the M¨¦daille K.Y. Lo Medal of the EIC,
and was the subject of a TIME Magazine feature article.
He was also awarded the unique Special University Prize in Applied Science of the University of British Columbia,
and the J. Gordin Kaplan Award for Research of the University of Alberta.

\end{IEEEbiography}

\begin{IEEEbiography}[{\includegraphics[width=1in,height=1.25in,clip,keepaspectratio]{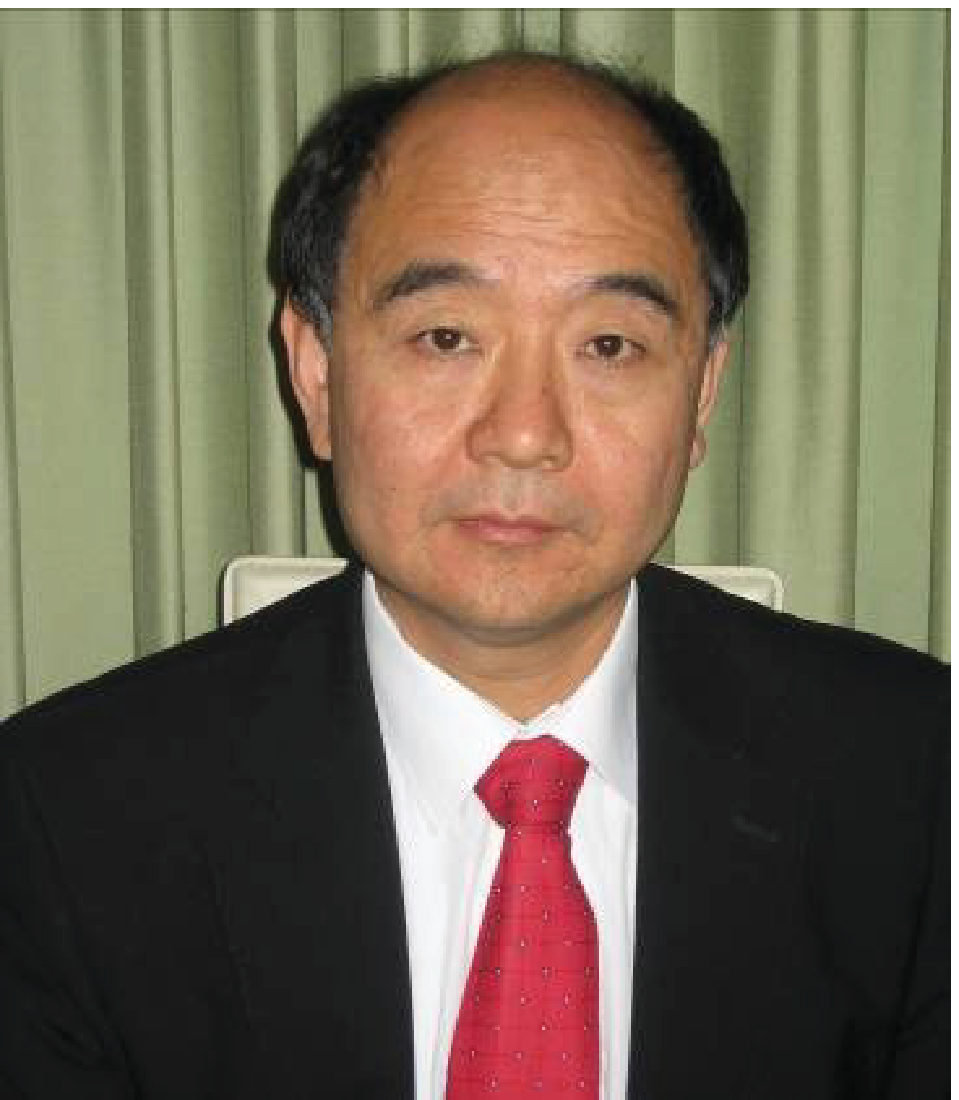}}]{Y. Jay Guo}

(F'14) received a Bachelor Degree and a Master Degree from Xidian University in 1982 and 1984, respectively,
and a PhD Degree from Xi'an Jiaotong University in 1987, all in China.
He is a Fellow of the Australian Academy of Engineering and Technology,
a Fellow of IEEE and a Fellow of IET, and a member of the College of Experts of Australian Research Council (ARC).
He has won a number of most prestigious Australian national awards, and was named one of the most influential engineers in Australia in 2014 and 2015.
His research interest includes antennas, mm-wave and THz communications and sensing systems as well as big data.
He has published over 300 research papers and holds 22 patents in antennas and wireless systems.

Prof Guo is a Distinguished Professor and the founding Director of Global Big Data Technologies Centre at the University of Technology Sydney (UTS), Australia.
Prior to this appointment in 2014, he served as a Director in CSIRO for over nine years, directing a number of ICT research portfolios.
Before joining CSIRO, he held various senior leadership positions in Fujitsu, Siemens and NEC in the U.K.

Prof Guo has chaired numerous international conferences.
He was the International Advisory Committee Chair of IEEE VTC2017, General Chair of ISAP2015,
iWAT2014 and WPMC'2014, and TPC Chair of 2010 IEEE WCNC, and 2012 and 2007 IEEE ISCIT.
He served as Guest Editor of special issues on ``Antennas for Satellite Communications''
and  ``Antennas and Propagation Aspects of 60-90GHz Wireless Communications,'' both in IEEE Transactions on Antennas and Propagation,
Special Issue on ``Communications Challenges and Dynamics for Unmanned Autonomous Vehicles'',
IEEE Journal on Selected Areas in Communications (JSAC), and Special Issue on ``5G for Mission Critical Machine Communications'', IEEE Network Magazine.

\end{IEEEbiography}

\end{document}